\documentstyle[epsf,12pt,citesort]{article}
\newcommand{\be}{\begin{equation}}
\newcommand{\ee}{\end{equation}}
\newcommand{\bea}{\begin{eqnarray}}
\newcommand{\eea}{\end{eqnarray}}

\newcommand{\hoch}[1]{$\, ^{#1}$}

\newcommand{\fft}[2]{{{#1} \over {#2}}}
\newcommand{\ft}[2]{{\textstyle{{\scriptstyle {#1}}\over {\scriptstyle {#2}}}}}

\newcommand{\dalemb}[2]{{\vbox{\hrule height .#2pt
        \hbox{\vrule width.#2pt height#1pt \kern#1pt
                \vrule width.#2pt}
        \hrule height.#2pt}}}
\newcommand{\square}{\mathord{\hbox{\hskip1pt}\dalemb{6.8}{7}\hbox{\hskip1pt}}}

\def\pa{\partial}

\makeatletter \@addtoreset{equation}{section} \makeatother

\def\H{{\cal H}}
\def\L{{\cal L}}
\newtheorem{theorem}{Theorem}

\begin{document}

\begin{flushright}
\hfill{\bf hep-th/0301187}\\

\hfill{MCTP-03-01}\\
\hfill{PUPT-2073}

\end{flushright}

\begin{center}

{\Large {\bf On Horizons and Plane Waves}}

\vspace{15pt}
James T. Liu\hoch{1}, Leopoldo A. Pando Zayas\hoch{2} and
Diana Vaman\hoch{3}

\vspace{7pt}
\hoch{1,2} {\it Michigan Center for Theoretical Physics,
University of Michigan\\
Ann Arbor, MI 48109--1120}

\vspace{7pt}
\hoch{2} {\it School of Natural Sciences,
Institute for Advanced Study\\
Princeton, NJ 08540}

\vspace{7pt}
\hoch{1,3} {\it Department of Physics,
Princeton University\\
Princeton, NJ 08544}

\vspace{2cm}
\underline{ABSTRACT}
\end{center}
We investigate the possibility of having an event horizon within several
classes of metrics that asymptote to the maximally supersymmetric IIB
plane wave. We show that the presence of a null Killing vector (not
necessarily covariantly constant) implies an effective separation of the
Einstein equations into a standard and a wave component.  This feature
may be used to generate new supergravity solutions asymptotic to the
maximally supersymmetric IIB plane wave, starting from standard seed
solutions such as branes or intersecting branes in flat space.  We find
that in many cases it is possible to preserve the extremal horizon of
the seed solution.  On the other hand, non-extremal deformations of the
plane wave solution result in naked singularities.  More generally, 
we prove a no-go theorem against the existence of horizons for backgrounds
with a null Killing vector and which contain at most null matter fields.
Further attempts at turning on a nonzero Hawking temperature by
introducing additional matter have proven unsuccessful.
This suggests that one must remove the null Killing vector in order to
obtain a horizon.  We provide a perturbative argument indicating that
this is in fact possible.

\newpage
\section{Introduction}

Black holes play a special role in our understanding of nature: Arising
as classical solutions to  gravity equations they pose challenging
questions at the quantum level. In the framework of general relativity
the best understood situations pertain to asymptotically flat black
holes. Motivated by the AdS/CFT correspondence \cite{ads} interest has
been directed to the study of black holes that asymptote to AdS. These
black holes have properties that fundamentally distinguish them from
asymptotically flat black holes, like the possibility of a positive
specific heat. Some of the properties of the simplest black hole
in this class --- the Schwarzschild-AdS black hole --- were  studied in
\cite{hawking}.  Most of the properties of asymptotically AdS black
holes have received beautiful interpretations in the framework of
the AdS/CFT correspondence. In particular, Witten argued that the
Hawking-Page phase transition is the supergravity dual effect to the
confinement/deconfinement phase transition for gauge theories on the
sphere \cite{witten}. Other properties of black holes asymptoting to AdS
have been discussed in the literature (see for example \cite{clifford}).

Recently Berenstein, Maldacena and Nastase (BMN) \cite{bmn} have
proposed a correspondence between a limit of the  AdS$_5\times S^5$
IIB background and a subsector of operators of ${\cal N}=4$ Super
Yang-Mills. The limit on the IIB supergravity side is the maximally
supersymmetric plane wave background with constant null RR five-form
(the BFHP solution) \cite{blau}
\bea
\label{bfhp}
ds^2&=&-4 dx^+dx^- -\ft1{16}\mu^2 z_iz^i (dx^+)^2+ dz_idz^i, \nonumber \\
F_{+1234}&=&F_{+5678}\,\,\,=\,\,\,\mu.
\eea
This solution is the analogue of the maximally supersymmetric plane wave
of 11 dimensional supergravity \cite{kg}.

One of the natural questions that arises from the BMN construction relates
to the nonzero temperature deformation of the correspondence, much in the
same way that the  AdS/CFT correspondence extends to nonzero temperature
\cite{witten}. On the string theory side some progress in formulating the
correspondence has been achieved. Exploiting the fact that the maximally
supersymmetric IIB plane wave is exactly solvable in the light-cone gauge
\cite{metsaev,metse},  the thermal partition function has been calculated
and shown to display a Hagedorn temperature depending on
the parameter $\mu$ above \cite{thermal}. This nontrivial dependence of the Hagedorn
temperature makes the possibility of a phase transition even more
tantalizing. Many natural questions however, remained unanswered. In
particular, the structure of the analogue of the Schwarzschild-AdS black
hole in the context of plane waves is not known. More generally: What are
the properties of black holes that asymptote to the homogeneous plane
wave spacetime?

The question of the Schwarzschild black hole asymptoting to
the BFHP plane wave has been addressed from different angles
\cite{cobi,donald,hr1,hr2}. In \cite{cobi,donald} it was made explicit
that the answer can not come from a Penrose limit of the Schwarzschild-AdS
black hole. This was further elaborated upon in \cite{hr1} where it
was also shown  that the existence of a covariantly constant
null Killing vector is an obstruction to having an event horizon. In
\cite{hr2} a further study of asymptotically plane wave spacetimes was
carried out and some solutions were presented using the Garfinkle-Vachaspati
method \cite{garfinkle} (see also \cite{myers}).

In this paper we investigate various possibilities for introducing event
horizons in geometries that asymptote to the BFHP plane wave. We
consider two fundamentally different classes of solutions: extremal and
nonextremal. More precisely, we describe solutions with
zero  Hawking temperature and attempt to generalize them to solutions
with nonzero Hawking temperature.

To construct the first class of solutions, that is, zero temperature
black objects containing an event horizon and asymptoting to the BFHP
plane wave, we make use of a separation property of backgrounds with
a null Killing vector.  Namely, we consider metrics that generalize the
BFHP plane wave while maintaining a null Killing vector:
\begin{equation}
\label{gmetric}
ds^2=e^{2A}[-4dx^+dx^-+{\cal H}dx^{+\,2}]+\sum_ie^{2B_i}dz_i^2,
\end{equation}
where $A(\vec z\,)$, $B_i(\vec z\,)$ and ${\cal H}(\vec z\,)$ are in
general functions of all transverse coordinates $\vec z$.
A crucial observation that may be used to establish various properties
is the fact that the Ricci tensor decomposes \cite{garfinkle}
into standard and plane wave components:
\begin{equation}
R_{MN}=R_{MN}^{(0)}-\ft12\delta_M^+\delta_N^+e^{2A}\sum_ie^{-2B_i}
[\partial_i\partial_i{\cal H}+\partial_i{\cal H}\partial_i{\cal G}_i],
\label{ricci}
\end{equation}
where $R_{MN}^{(0)}$ is computed from (\ref{gmetric}) when ${\cal H}=0$ and
${\cal G}_i=2A+\sum_jB_j-2B_i$.

A very natural way to exploit the property (\ref{ricci}) is to use it as
a solution generating mechanism. Given that only the $R_{++}$ component
of the Ricci tensor is modified, this
allows for the possibility of ``superposing'' a plane wave on top of a
given solution. Namely, finding a suitable matter contribution $\delta
T_{++}$ we can generalize a given supergravity solution with $\H=0$ to a
new solution with $\H\ne 0$. Imposing that asymptotically $A,B
\longrightarrow 0$ and that $\H$ approaches the form of (\ref{bfhp}) we
construct solutions that asymptote to the BFHP plane wave. 

Along the lines of the solution generating point of view we can reinterpret
the BFHP solution as the  $\H$-deformation of flat space. Under this scope
we review the construction of D branes and more importantly, we obtain new
solutions corresponding to $\H$-deforming intersecting branes. We find
it possible in some cases to arrange for the $\H$-deformation to keep the
regular horizon
characteristic of the seed solutions while changing the solutions
asymptotically to that of BFHP for large $\vec{z}$. This way we present a
class of solutions that asymptote to the BFHP plane wave and have
regular event horizons.

We then attempt to further generalize such solutions to include an event
horizon corresponding to nonzero Hawking temperature. We study possible
generalizations under various symmetry requirements. In particular we
consider $SO(8)$ and $SO(4)\times SO(4)$ generalizations of the BFHP
solution. We find that in the presence of a null Killing vector such
generalizations yield singular solutions. Using the separation
property of the Ricci tensor (\ref{ricci}), we show that generalizations
of the BFHP solution keeping a null Killing vector and with arbitrary
splitting of the transverse directions can not lead to a smooth
horizon. Following the expectations raised in the zero temperature case
we go on to consider nonzero Hawking temperature generalizations of the
$\H$-deformed branes and intersecting brane solutions. We find however,
that all such generalizations yield singular event horizons.

Having identified the main obstruction to having a nonzero Hawking
temperature event horizon as the presence of a null Killing vector,
we consider a perturbative argument in favor of the existence of such
solutions once the null Killing vector is removed. Namely, we consider
the black string in ten dimensions and discuss the conditions for turning
on a RR 5-form. We find, neglecting backreaction, compatibility between
the Bianchi identity and the asymptotic conditions. Moreover, we show
explicitly that the backreaction is under control.

Because of our interest in identifying regular event horizons, we begin
in section 2 with an analysis of the horizon structure in generalized
plane wave backgrounds.  In section 3, we exploit the separation property
of the Ricci tensor to construct new multiple charge solutions
with zero temperature, including a deformation of the well-known D1-D5
system.  Then, in section 4, we investigate the possibility of finding
solutions with non-zero temperature.  We find that the presence of a
null Killing vector presents an obstruction to the existence of such
solutions.  This leads us to a no-go theorem proved in section 5.  In
section 6, we consider turning on additional sources (while maintaining
a null Killing vector), but show that such examples are unable to
overcome the no-go theorem.  Finally, we conclude in section 7 by
demonstrating (at least perturbatively) how a horizon may be obtained
by relaxing the null Killing vector condition.

\section{Asymptotic time and event horizons in plane waves}
\label{sec:n2}

Since our main endeavor is dedicated to finding supergravity solutions
that share the same asymptotics as the maximally supersymmetric IIB plane
wave and that have a regular event horizon, we devote this section to
examining the criteria for the existence of an event horizon. We also discuss 
how  different asymptotics might change the analysis of
the event horizon. First, a cautionary word about event horizons in
asymptotically non flat spaces is in order. A rigorous definition
of event horizons in the case of asymptotically flat black holes is
given, for example, in \cite{wald} ({\it pp.}~299--300). Namely, a black
hole $B$ is defined as a region of a strongly asymptotically predictable
spacetime $(M,g_{ab})$ given by $B=[M-J^-({\cal J}^+)]$. This definition
is equivalent under very mild topological conditions to $B\cap J^-({\cal
J}^+)=\emptyset$, where $J^-({\cal J}^+)$ denotes the causal past
of the future null infinity.  However, such a
definition is lacking for general asymptotics, in part because of the
difficulties in identifying ${\cal J}^+$. (Discussions of the causal
structure in spaces asymptoting to plane wave have appeared in
\cite{horatiu,donross,hr3}.) The need for finding criteria
for an event
horizon without requiring knowledge of the global development of a
spacetime was addressed partially in \cite{wald} ({\it
pp.}~308--312). Along those lines we will, as a working definition,
declare a black hole to be a region causally disconnected from asymptotic
infinity.

Our analysis, however, uses asymptotic infinity rather indirectly by
replacing it by a suitable causally connected region $R$. Thus, we only need
to show that this region $R$, which is causally connected to asymptotic
infinity, is causally disconnected from the black hole region. Therefore,
our analysis
boils down to the study of the proper (affine)  time and the time of
propagation of signals along timelike and null geodesics. Namely, we
require the proper (affine) time  to be finite so that the spacetime
can be extended beyond that point but we require the (coordinate) time
traveled along a timelike or null geodesic to be infinite so that causal
communication is impossible.  Note that, by coordinate time, we mean a
suitable time coordinate conjugate to a timelike Killing vector.  Up to
finite time dilation, this would be the time measured by a static
observer in the region $R$.

Of course, it may be the case that there is not a unique timelike
Killing vector.  In this case, different choices may be made for
coordinate time, corresponding to different observers boosted with
regard to each other.  For finite boosts, all such observers are in
causal contact.  Thus if any single observer ({\it i.e.} choice of time
coordinate) can communicate with the purported black hole region, then
a horizon cannot exist and the ``black hole'' would instead be a
singularity of some sort (assuming a curvature singularity develops).
As a result, our procedure is clear.  To demonstrate the existence of a
horizon, we must show that all possible choices of coordinate time yield
an infinite $\Delta t$ along the geodesic.  In practice, however, it is
often sufficient to consider only a handful of choices, as will be made
clear below.

For the geometries we are interested in, it is sufficient to discuss only
metrics which admit a null Killing vector.  Let us begin by considering a
generalization of the plane wave metric of the form:
\be
ds^2=e^{2A(r)}[-4 dx^+ dx^- +{\cal H}dx^{+\,2}] + e^{2B(r)} dr^2 + \cdots,
\label{eq:gpwm}
\ee
where ellipses signify other coordinates which are unimportant for the
present purpose. In addition, let us assume that there is an event horizon
at some finite distance $r_0$.  Then one can study geodesics crossing the
horizon and estimate the affine time and the time as measured by an
observer in causal contact with infinity.
To investigate the conditions under which an event horizon might exist,
we consider timelike and null radial geodesics in this geometry. The geodesic
equations can be obtained from the following effective Lagrangian:
\be
{\cal L} = e^{2A(r)}[-4\dot{x}^+ \dot{x}^- + {\cal H}\dot{x}^{+\,2}] + e^{2B(r)}
\dot{r}^2,
\ee
where dots indicate differentiation with respect to the proper time.
Since the above Lagrangian does not depend explicitly on $x^+$ and
$x^-$, we have two integrals of motion, $E_+$ and $E_-$, which allow to
solve for the motion along $x^+$ and $x^-$ as
\be
\dot{x}^+=e^{-2A} E_-, \qquad \dot{x}^-={1\over 2}e^{-2A}(\H E_--E_+).
\ee
We set  ${\cal L}=-1,0$ for timelike and null geodesics respectively
and solve for motion along the radial coordinate
\be
\dot{r}^2 =e^{-2B}\bigg[e^{-2A}E_-(\H E_--2E_+)+\L\bigg].
\ee
By integrating this equation we find the proper time needed to travel
between two radial positions $r_{\rm in}$ and $r_{\rm fin}$ along an
incoming geodesic
\be
\label{proper}
\Delta \tau =\int\limits_{r_{\rm fin}}^{r_{\rm in}}dr\,
e^{B}\bigg[e^{-2A}E_-(\H E_--2E_+)+\L\bigg]^{-1/2}.
\label{Deltau}
\ee
Similarly we have the following formal variations of $x^+$ and $x^-$:
\begin{eqnarray}
\label{times}
\Delta x^+&=& E_-\int\limits_{r_{\rm fin}}^{r_{\rm in}}e^{-2A+B}
\bigg[e^{-2A}E_-(\H E_--2E_+)+\L\bigg]^{-1/2}\, dr,\nonumber\\
\Delta x^-&=&{1\over 2} \int\limits_{r_{\rm fin}}^{r_{\rm in}}
e^{-2A+B}(\H E_--E_+)\bigg[e^{-2A}E_-(\H
E_--2E_+)+\L\bigg]^{-1/2}\, dr.\label{deltapm}
\end{eqnarray}

To identify an asymptotic time coordinate for the metric (\ref{eq:gpwm}),
we note that it admits two Killing vectors, $\partial/\partial x^+$ and
$\partial/\partial x^-$.  Thus we may choose to measure time as
\begin{equation}
\Delta t = c^+\Delta x^++c^-\Delta x^-
\label{eq:deltdef}
\end{equation}
(with constant $c^+$ and $c^-$), so long as $\Delta t$ is a timelike
interval.  This results in the condition $c^+(c^+\H-4c^-)<0$.  So long
as we are interested in solutions that asymptote to the maximally
supersymmetric IIB plane wave, then
\be
\H\stackrel{r\rightarrow \infty}{\longrightarrow}  -\ft1{16}\mu^2 r^2,
\ee
so that $\partial/\partial x^+$ by itself is a timelike Killing vector.
In this case $x^+$ can be taken as a natural asymptotic time coordinate,
and we may set $c^-=0$ in (\ref{eq:deltdef}).  However, for
$\partial/\partial x^+$ either null or spacelike, we cannot set $c^-=0$,
and hence must resort to $x^-$ (or some combination of $x^+$ and $x^-$)
as an asymptotic time coordinate.  We will make a distinction between
these two possibilities below.

Since we are interested in motion near the purported horizon at $r_0$,
a generic behavior for the warp factors can be modeled by 
\be
e^{2A}\sim (r-r_0)^{2a},\qquad e^{2B}\sim (r-r_0)^{2b}.
\label{horbehav}
\ee
The behavior of the function $\H$ can be a little more involved. Let us
assume a power-like behavior:
\be
\H\sim (r-r_0)^{2h} {\rm sgn}\,\H.
\label{hhorbehav}
\ee
For (\ref{proper}) and (\ref{times}) to make sense we need the square
root to be real. We first consider the case $\H\to const.$ at the
horizon, {\it i.e.} $h\ge 0$. Taking $E_+<0$ we find
\be
\Delta\tau \sim\int_0^\epsilon \,dx\, x^{a+b}, 
\qquad \Delta x^+ \sim \int_0^\epsilon \,dx\, x^{b-a},
\qquad \Delta x^- \sim \int_0^\epsilon \,dx\, x^{b-a}.
\ee
In this case, regardless of the choice of time coordinate,
(\ref{eq:deltdef}),
this means that to have a horizon we need $a+b>-1$ and $b-a\le -1$. On
the other hand, if $\H$ grows $(h<0)$, we need ${\rm sgn}\,\H=1$
(otherwise we would encounter a repulsive singularity much like in the
negative mass Schwarzschild black hole). In this case  
\be
\Delta \tau\sim\int_0^\epsilon \,dx \,x^{a+b+|h|},\qquad 
\Delta x^+ \sim \int_0^\epsilon\,dx\,x^{b-a+|h|},\qquad
\Delta x^- \sim \int_0^\epsilon\,dx\,x^{b-a-|h|}.
\label{eq:tauxpxm}
\ee
Here care must be taken in relating asymptotic time to $x^+$ and $x^-$.
Note that $\Delta x^-$ is always strictly more divergent than $\Delta
x^+$ since $|h|>0$.  Hence the shortest attainable $\Delta t$ is given
by using $x^+$ as asymptotic time, which is only possible for
$\partial/\partial x^+$ timelike.  Since all other observers would
measure a longer time interval, we only need to check that $\Delta x^+$
diverges.  As a result, the horizon conditions are $a+b+|h|>-1$ and
$b-a+|h|\le -1$.  On the other hand, if $\partial/\partial x^+$ were
null or spacelike, $\Delta t$ would always include a non-vanishing
contribution from $\Delta x^-$.  In this case, the second horizon
condition is modified to $b-a-|h|\le-1$, which is a weaker condition
than that for $\partial/\partial x^+$ timelike.  Finally, let us note
that for a logarithmic behavior, $\H\sim |\log(r-r_0)|\,{\rm sgn}\H$,
the situation is effectively identical to the $h\ge 0$ case. The only
novelty is that  ${\rm sgn}\,\H=-1$ leads to a repulsive, negative mass
type of behavior. 

To summarize, for solutions that asymptote to the maximally
supersymmetric IIB plane wave, in which case $\partial/\partial x^+$ is
timelike, the conditions for the existence of a horizon are
\bea
&a+b>-1,\qquad a-b\ge 1\qquad&{\rm if}~~
h\geq 0;\nonumber\\
&a+b+|h|>-1,\qquad a-b\ge1+|h|\qquad&{\rm if}~~ h<0
~~{\rm and}~~{\rm sgn}\,\H=+1;\nonumber\\
&a+b>-1,\qquad a-b\ge 1\qquad&{\rm if}~~
\H\sim |\log(r-r_0)|,
\label{eq:hcabh}
\eea
where $a,b$ and $h$ have been defined in (\ref{horbehav}) and
(\ref{hhorbehav}), respectively.  Note that the second line is stronger
than the first line.  Hence if no horizon is possible for $h\ge0$,
neither can it be possible for $h<0$.  

More generally, one can state that the criteria (\ref{eq:hcabh})
apply for any background whose asymptotics are those of a homogeneous
plane wave, {\it i.e.} $H\stackrel{r\rightarrow \infty}{\longrightarrow} 
-\mu^2_{ij} x^i x^j$.
Since $R_{++}$ is positive by the weak energy condition, and $R_{++}$
is the sum of the eigenvalues of $\mu^2_{ij}$, this implies that
$\partial/\partial x^+$ is asymptotically a time-like Killing vector;
therefore all results previously derived are valid.  On the other hand,
for the case where $\partial/\partial x^+$ is null or spacelike, our
above analysis still holds, provided the second condition in (\ref{eq:hcabh})
is replaced by
\begin{eqnarray}
&a+b+|h|>-1,\qquad a-b\ge1-|h|\qquad&{\rm if}~~ h<0
~~{\rm and}~~{\rm sgn}\,\H=+1.
\label{eq:hcabh2}
\end{eqnarray}
This condition is now weaker than for the $h\ge0$ case.

The set of conditions for the existence of a horizon defined as the boundary 
of a region causally disconnected from infinity, (\ref{eq:hcabh}), can be equally
well derived from a local analysis of the horizon. In the latter case we define
the horizon as the boundary of a region of no escape; the idea behind this 
local definition is that gravity bends light, and  
a massive enough object could ultimately trap the light wavefronts.
If this were to happen, then the outgoing light wave fronts would have a decreasing
area, finally reaching the horizon whereupon the light rays would be instantaneously 
parallel. Since the rate of change in the area is given by the 
scalar expansion $\theta$, we turn to a 
computation of the expansions associated with the two null vectors 
pointing in the directions of the incoming and outgoing light wavefronts.
For instance, the expansion associated with the the vector tangent to 
the null geodesic $k^\mu=(\dot x^+, \dot x^-, \dot r,\vec 0\,)$ is 
$\theta_k=\widehat{D_\mu k^\mu}$ where the hat denotes the implicit projection
onto the $d-2$ vector space orthogonal to the null vectors. The
other null vector is normalized by $l^\mu k_\mu=-1$, and since it 
corresponds to the outgoing light wavefront it has a non-vanishing radial 
component. 
The trapping horizon is the closure of a  
hypersurface foliated by marginal surfaces, on which $\theta_l$ vanishes
and $\theta_k$ is non-vanishing and negative (see for example
\cite{hayward}). 
Therefore in the trapped region
we have $\theta_k \theta_l>0$. Without going into further detail, one 
obtains the weaker of the two conditions in (\ref{eq:hcabh})
from the requirement that the second expansion is non-zero and negative
at the horizon, while the strongest
condition (the fact that infinity and the interior of the horizon are causally
disconnected) is equivalent with the fact that the first expansion (the
$\theta_l$ one) vanishes.

\subsection{Other asymptotic times}
For spacetimes asymptotic to the plane wave, we have been able to
identify the asymptotic time with $x^+$. This is, however, not the case
for various examples of known solutions which have a null Killing vector
and a regular horizon. To clarify this point and to warn the reader that
the horizon criteria must be applied with care, we will present two
examples where (\ref{eq:hcabh}) is not directly applicable. 

The first example is the five dimensional three charge black hole solution.
When lifted to ten dimensions it is mapped into a D1-D5 system
(in the string frame)
\be
ds^2 = (H_1 H_5)^{-1/2} (-dx^+ dx^- + K dx^+ {}^2)+(H_1 H_5)^{1/2}
d{\vec x}^{\,2}_{4} + H_1^{1/2}H_5^{-1/2} d{\vec y}^{\,2}_{4},
\label{eq:3c5dbh}
\ee
where the three charges of the 5d black hole correspond to the number of
D1 branes, $Q_1$, the number of D5 branes, $Q_5$, and the momentum along
the compactified 9-th direction, $N$:
\bea
H_1(r)=1+\frac {c_1Q_1}{r^2},\qquad H_5(r)=1+\frac {c_5Q_5}{r^2},\qquad
K(r)=\frac {c_wN}{r^2},
\eea
and where $r^2=|\vec{x}\,|^2$.
Given that this solution asymptotes to Minkowski flat space, we may
choose
\be
t=\frac{x^+ +x^-}2,\qquad x_9=\frac{x^- -x^+}2,
\ee
(or a similar but boosted frame).  The clock of the observer at infinity
measures time in the coordinate $t$ rather than $x^+$, as would be
natural for the case of solutions asymptoting to BFHP.  Using the two
integrals of motion
\be
E=(H_1 H_5)^{-1/2} ((1-K)\dot t+K\dot x_9),\qquad
E_9=(H_1 H_5)^{-1/2}((1+K)\dot x_9-K\dot t),
\ee
the radial timelike geodesic, in the vicinity of $r=0$, is given by
\be
-1=\frac 1{r^2}\bigg[ -(1+K)E^2+(1-K)E_9^2+2KE E_9\bigg]+\frac{\dot r^2}{r^2}.
\ee
Hence the proper time spent nearby the horizon is finite:
\be
\int d\tau\sim \int r\,dr,
\ee
while, from the point of view of the observer at infinity, the time elapsed
is infinite
\be
\int dt\sim \int \frac{dr}{r}.
\ee
These results are a particular case of (\ref{eq:tauxpxm}), where we must
consider that near $r=0$ it is $\Delta x^-$ which provides the leading
behavior of $\Delta t$.

The area of the horizon is finite, and the entropy is solely determined
by the product of the three charges, $S={A_H}/{4G_N}=2\pi\sqrt{NQ_1Q_5}$.
Note that for this solution
the norm of the Killing vector $\xi=\partial/\partial t$ is vanishing 
on the ergosphere,
$1-K(r_E)=0$, and is 
positive for $0 \leq r \leq r_E$. A similar situation
was encountered for  the four dimensional charged Kerr black hole. This
means only that to pass through the stationary surface at $r_E$
following a timelike radial geodesic, one needs motion in the $x_9$
direction as well; {\it i.e.} setting $\dot x_9=0$ would be inconsistent.

The second example that we want to discuss is the 5d black string
obtained in \cite{5dstring} (see also \cite{stringmore,myers}).
This five dimensional solution, whose metric in the string frame is given by
\be
ds^2= \frac fh dx^{+\,2}+\frac 2h dx^+ dx^- + kl(dr^2+ r^2 d\Omega_3^2),
\ee
is characterized by four charges
\bea
f(r)&=&1+\frac {Q_1}r,\qquad h(r)=1+\frac {Q_2}r,\nonumber\\
k(r)&=&1+\frac{P_1}r,\qquad l(r)=1+\frac {P_2}r.
\eea
The asymptotic metric is Minkowski, and the asymptotic time is naturally
identified as
\be
t=x^-,\qquad x_5=x^+ + x^-
\ee
(again up to a boost).
The event horizon is located at $r=0$, as we can convince ourselves by
estimating the proper time elapsed as the free falling particle approaches
the horizon
\be
-1=-\frac{E_-^2}{r^2}+2 \frac{E_- E_+}{r}+\frac {\dot r^2}{r^2},
\ee
where $E_-, E_+$ are integrals of motion associated with the
$\partial/\partial x^-, \partial/\partial x^+$ isometries.  Thus the
proper time is finite, $\int d\tau \sim \int dr$, while the observer
at infinity measures an infinite time, $\int dt\sim\int r^{-2}dr$.
As before, there is a stationary surface where $g_{tt}$ vanishes, and
to reach the horizon at $r=0$ the free falling particles must move in
the $x_5$ direction as well.

\section{Solution generation with a null Killing vector}
\label{sec:n3}

Having examined the criteria for the existence of a regular event
horizon in a generalized plane wave metric, we now turn to some
examples.  Since  a covariantly constant
null Killing vector precludes the existence of a regular horizon, here
we relax the covariant constancy but maintain a null Killing vector.  In
this case, we may apply the observation of Garfinkle and Vachaspati
\cite{garfinkle} (see also \cite{myers}) that such a metric with a null
Killing direction gives rise to a separation property of the Ricci
tensor, (\ref{ricci}).  As indicated below, this property may be
exploited to construct new wave solutions, given an initial ``seed''
solution.

\subsection{The Ricci tensor in the presence of a null Killing vector}

While the separation property of \cite{garfinkle,myers} is more
general than the one considered here, for the applications considered in
this paper, it is sufficient to take a generic metric of the form
\begin{equation}
\label{generalmetric}
ds^2=e^{2A}[-4dx^+dx^-+{\cal H}dx^{+\,2}]+\sum_ie^{2B_i}dz_i^2,
\label{eq:glcmet}
\end{equation}
where $A(\vec z\,)$, $B_i(\vec z\,)$ and ${\cal H}(\vec z\,)$ are in
general functions of all transverse coordinates $\vec z$.  Note that
this form of the metric allows us to subsequently split the $z_i$
directions into additional longitudinal, relative-transverse and
overall-transverse directions by taking the functions to be independent
of some of the $z_i$'s.

For this metric, the non-vanishing Christoffel symbols are
\begin{eqnarray}
&\Gamma^+{}_{+i}=\partial_iA,\qquad\Gamma^-{}_{-i}=\partial_iA,\qquad
\Gamma^-{}_{+i}=-\ft14\partial_i{\cal H},&\nonumber\\
&\Gamma^i{}_{++}=-g_{++}e^{-2B_i}(\partial_iA+\ft12\partial_i\log{\cal
H}),\qquad
\Gamma^i{}_{+-}=-g_{+-}e^{-2B_i}\partial_iA,&\nonumber\\
&\Gamma^i{}_{jk}=\delta_{ij}\partial_kB_i+\delta_{ik}\partial_jB_i
-e^{2B_j-2B_i}\delta_{jk}\partial_iB_j.&
\end{eqnarray}
In this expression and below, indices are not summed over unless
explicitly indicated.  A straightforward computation yields for the
non-trivial components of the Ricci tensor
\begin{eqnarray}
R_{+-}&=&-g_{+-}\sum_ie^{-2B_i}[\partial_i\partial_iA+\partial_iA\partial_i
{\cal G}_i],\nonumber\\
R_{++}&=&-g_{++}\sum_ie^{-2B_i}[\partial_i\partial_iA+\partial_iA\partial_i
{\cal G}_i]
-\ft12e^{2A}\sum_ie^{-2B_i}[\partial_i\partial_i{\cal H}
+\partial_i{\cal H}\partial_i{\cal G}_i],\nonumber\\
R_{ij}&=&-\ft12\partial_i\partial_j({\cal G}_i+{\cal G}_j)
-2\partial_iA\partial_jA+2(\partial_iA\partial_jB_i+\partial_jA\partial_iB_j)
\nonumber\\
&&+\sum_k(\partial_iB_j\partial_jB_k+\partial_jB_i\partial_iB_k)
-\sum_k\partial_iB_k\partial_jB_k-2\partial_iB_j\partial_jB_i\nonumber\\
&&-g_{ij}\sum_ke^{-2B_k}[\partial_k\partial_kB_i+\partial_kB_i\partial_k
{\cal G}_k],
\label{eq:glcri}
\end{eqnarray}
where ${\cal G}_i=2A+\sum_jB_j-2B_i$.  Note, in particular, that ${\cal
H}$ only appears in $R_{++}$.  As a result, this allows us to write
\begin{equation}
R_{MN}=R_{MN}^{(0)}-\ft12\delta_M^+\delta_N^+e^{2A}\sum_ie^{-2B_i}
[\partial_i\partial_i{\cal H}+\partial_i{\cal H}\partial_i{\cal G}_i],
\label{eq:rmnsplit}
\end{equation}
where $R_{MN}^{(0)}$ is computed from (\ref{eq:glcmet}) when ${\cal H}=0$.
Note that, up to a suitable coordinate transformation, $R_{MN}^{(0)}$ is
simply the Ricci tensor for the ``conventional'' metric
\begin{equation}
ds^2=e^{2A}[-dt^2+dw^2]+\sum_ie^{2B_i}dz_i^2.
\end{equation}
A particular case of (\ref{generalmetric}) that we will often be
interested in are metrics with a split transverse
space, of the form
\begin{equation}
ds^2=e^{2A}[-4dx^+dx^-+{\cal H}dx^{+\,2}]+e^{2B}d\vec z_d^{\,2} +
e^{2C}d\vec y_n^{\,2}.
\end{equation}
The Ricci components for this metric may be obtained from
(\ref{eq:glcri}) by taking $B_i=B$ for $i=1,2,\ldots,d$ and $B_a=C$ for
$a=d+1,d+2,\ldots,d+n$ and furthermore by demanding that $A$, $B$ and $C$
are independent of $z_a$.  The resulting Ricci components are
\begin{eqnarray}
R_{+-}&=&-g_{+-}e^{-2B}[\square_d A+\partial_iA\partial_i {\cal G}],\nonumber\\
R_{++}&=&-g_{++}e^{-2B}[\square_d A+\partial_iA\partial_i {\cal G}]
-\ft12e^{2A-2B}[\square_d{\cal H}+e^{2B-2C}\widetilde{\square}_n{\cal H}
+\partial_i{\cal H}\partial_i{\cal G}],\nonumber\\
R_{ab}&=&-g_{ab}e^{-2B}[\square_d C+\partial_i C\partial_i{\cal
G}],\nonumber\\
R_{ij}&=&-\partial_i\partial_j{\cal G}-2\partial_iA\partial_jA
+2(\partial_iA\partial_jB+\partial_jA\partial_iB)+(d-2)\partial_iB\partial_jB
\nonumber\\
&&+n(\partial_iB\partial_jC+\partial_jB\partial_iC
-\partial_iC\partial_jC)
-g_{ij}e^{-2B}[\square_d B+\partial_kB\partial_k {\cal G}],
\label{eq:dnsplit}
\end{eqnarray}
where ${\cal G}=2A+(d-2)B+nC$ and $\square_d\equiv\sum_{i=1}^d
\partial_i\partial_i$ and
$\widetilde{\square}_n=\sum_{a=d+1}^{d+n}\partial_a\partial_a$.  While
$\square_d$ is the flat transverse Laplacian, we note that the covariant
Laplacian
\begin{equation}
\widehat{\square}_d\equiv e^{-2B-{\cal G}}\partial_ie^{\cal{G}}\partial_i
=e^{-2B}[\square_d+(\partial_i{\cal G})\partial_i],
\label{eq:wtsq}
\end{equation}
is in fact the natural quantity that shows up in (\ref{eq:dnsplit}).  This
is also the combination appearing in (\ref{eq:rmnsplit}). The appearance of
the transverse Laplacian is the foundation of the Garfinkle-Vachaspati
method, which is based on finding solutions to $\square \H=0$
\cite{garfinkle,myers}. In this sense, the present solution generating
mechanism is based on finding solutions to the transverse Laplacian, but
with a source on the right hand side.

\subsection{Solution generating technique}
In this subsection we show explicitly how the previous separation of
$R_{MN}$ can
be used as a solution generating technique. We briefly review some known
solutions in the literature under this light, but also present new
solutions. Hopefully our presentation here will provide a unifying
framework for understanding a wide class of solutions that asymptote
to the BFHP solution.

The general construction states that given a supergravity solution of
$S_0$ we can generate a solution $S_{\H}$ where the metric in $S_{\H}$
asymptotes to the BFHP solution. Let us consider the Einstein equations
of motion for the seed solution $S_0$
\be
R_{MN}^{(0)}=T_{MN}^{(0)}.
\ee
This solution corresponds to $\H=0$. Due to the properties of the Ricci
tensor we see that it is possible to
construct a solution with $\H\ne 0$ granted we can provide a $\delta
T_{MN}$ that compensates the contribution of $\H$ to the Ricci
tensor. In particular we see that all we need to provide is a new field
whose only contribution to the stress energy tensor is of the form
$\delta T_{++}$. The candidate for such field is a null form, naturally
$F_5$. Note, however that this construction is not universal and one has
to guarantee that the introduction of the null 5-form be compatible
with the existing form fields in the seed solution. This nonuniversality
forces us to consider generating solutions on a case by case basis.

\subsubsection{The BFHP solution}
The simplest example of an $\H$-deformed solution is the BFHP
background itself, which can be seen as the $\H$-deformation
of ten dimensional Minkowski spacetime.
The maximally supersymmetric plane wave solution to IIB theory was constructed
in \cite{blau} as a supergravity background involving the metric
and RR 5-form flux.  In Brinkman coordinates, the solution has the form
\begin{eqnarray}
ds^2&=&-4dx^+dx^-+{\cal H}dx^{+\,2}+d\vec z\,^2,\nonumber\\
F_{5}&=&\mu\,dx^+\wedge(dz^1\wedge dz^2\wedge dz^3\wedge dz^4+
dz^5\wedge dz^6\wedge dz^7\wedge dz^8).
\label{eq:hpp}
\end{eqnarray}
Having ten dimensional Minkowski spacetime as the seed solution, its
non-deformed equations of motion are trivial. For the deformed solution
there is only one nontrivial equation:
\be
\square {\cal H}=-\mu^2.
\ee
Here it is seen that the 5-form
serves as a source to this equation (the $\delta T_{++}$ alluded to
previously), resulting in a general $SO(8)$
symmetric solution \cite{Cvetic:2002hi}
\begin{equation}
{\cal H}=c_0+\fft{{\cal Q}}{r^6}-\ft1{16}\mu^2r^2.
\label{eq:hppheqn}
\end{equation}
Setting ${\cal Q}=0$ results in the maximally supersymmetric plane wave
solution.

Many important features of this solution derive from the fact that the
plane wave metric admits a covariantly constant null Killing vector,
$\partial/\partial x^-$.  In particular, the only non-trivial Riemann
and Ricci components are
\begin{equation}
R_{+i+j}=-\ft12\partial_i\partial_j{\cal H}(\vec z\,),\qquad
R_{++}=-\ft12\square {\cal H}(\vec z\,).
\end{equation}
It is easy to show that any generalization of the plane wave metric
which preserve a covariantly constant null Killing vector does not
admit an event horizon.  The details of the argument were spelled out in
\cite{hr1}.

\subsubsection{$\H$-deformed 3-branes}
\label{sec:3.2.2}

We now turn to a generalization of D3 branes in a plane wave background.
These types of backgrounds have been described from the supergravity point
of view in, for example, \cite{bain,a1,kostas} and from the string theoretic
point of view in \cite{matthias,bain2,kostas2,matthias2}. Here we view these
solutions as $\H$-deformations of the standard D3-brane solution of IIB
supergravity.

There is however an ambiguity in the form of the new 5-form needed to
compensate for the introduction of nonzero $\H$.  We begin by presenting
the D3 $(+,-,1,1)$ solution.  The supergravity background in question is
\begin{eqnarray}
ds^2&=& H^{-1/2} [ -4 dx^+\, dx^- +{\cal H}(\vec z,x,y) \, dx^{+\,2}+ dx^2
+ dy^2]+ H^{1/2} d\vec z_6^{\,2}, \nonumber \\
F_5^{(0)}&=& 2\cdot(1+ *)\,dx^+ \wedge dx^- \wedge dx\wedge dy \wedge
dH^{-1},\nonumber\\
F_5^{(1)}&=& \mu \,dx^+\wedge ( dx\wedge dz^1\wedge dz^2\wedge dz^3
+ dy \wedge dz^4\wedge dz^5\wedge dz^6),
\label{eq:d3pm11}
\end{eqnarray}
where $F_5^{(0)}$ corresponds to the D3-brane source (giving rise to
$T_{MN}^{(0)}$) and $F_5^{(1)}$ is
the compensating null 5-form supporting the $\H$ deformation.
We find that, as explained previously, all equations are satisfied by
virtue of the seed solution with
\be
H=1+{L^4\over r^4},
\label{eq:d3good}
\ee
where the six dimensional radius is $r^2=\sum\limits_{i=1}^6z_i^2$.
We are left with only one equation involving $\H$ and the new source:
\be
\bigg[\square^{(6)} +H(\pa_x^2+ \pa_y^2)\bigg]\H=-\mu^2,
\ee
with general solution of the form
\be
{\cal H}=c_0 + {{\cal Q}\over r^4}
-\ft1{16}\mu^2 \left(\rho^2 + {L^4\over r^2}\right),
\ee
where $\rho^2=r^2+x^2+y^2$.
For ${\cal Q}=0$ we recover the solution presented in \cite{bain}. 

Let us analyze the possibility of a horizon for this solution from the
point of view of section \ref{sec:n2}. Exploring the possibility of a
horizon at $r=0$, we find $a=1$ and $b=-1$. However, the behavior of $\H$
(and hence the horizon analysis) depends on the charges present in the
solution.  For the standard D3 brane $(\mu=0, \, {\cal Q}=0)$ we find a
horizon as expected.  On the other hand, allowing ${\cal Q}\ne 0$ leads
to $h=-2$.  In this case, according to the criteria (\ref{eq:hcabh})
$r=0$ is no longer a horizon.  Finally, for ${\cal Q}=0$ but nonzero
$\mu$, we find that the dominant term in $\H$, using the language of
section \ref{sec:n2}, corresponds to $h<0$ and ${\rm sgn}\,\H=-1$ which
signals a repulsive singularity. Thus the $\H$-deformation of D3 branes
resulting in a generalization of the form D3 $(+,-,1,1)$ does not admit
a regular horizon.

Another possibility of consistently introducing a null 5-form supporting
nonzero $\H$  leads to a generalization of the D3 $(+,-,2,0)$ solution.
In this case, however, in order for the null 5-form to be compatible
with the initial $F_5^{(0)}$, the seed D3 brane must be delocalized
along two of its transverse directions.  The supergravity background is
then of the form
\begin{eqnarray}
ds^2&=& H^{-1/2} [ -4 dx^+\, dx^- +{\cal H}(\vec z,x,y,w,z) \, dx^{+\,2}+
dx^2 + dy^2]+ H^{1/2} [dw^2+dz^2 + d\vec z_4^{\,2}], \nonumber \\
F_5^{(0)}&=& 2\cdot(1+ *)\,dx^+ \wedge dx^- \wedge dx\wedge dy \wedge
dH^{-1},\nonumber\\
F_5^{(1)}&=& \mu \,dx^+\wedge ( dx\wedge dy\wedge dw\wedge dz
+ Hdz^1 \wedge dz^2\wedge dz^3\wedge dz^4),
\label{eq:d3pm20}
\end{eqnarray}
where again $F_5^{(0)}$ corresponds to the D3-brane source, while
$F_5^{(1)}$ is now turned on corresponding to the $(+,-,2,0)$ solution.
The D3 brane is delocalized along the $v$ and $w$ directions.  As a
result, instead of (\ref{eq:d3good}), the seed solution has
\be
H=1+{L^2\over r^2},
\ee
where $r^2=\sum\limits_{i=1}^4z_i^2$.  The equation for $\H$ is
\be
\bigg[\square^{(4)} +H(\pa_x^2+ \pa_y^2)+\partial_w^2+\partial_z^2\bigg]\H
=-\mu^2H,
\ee
which has a solution
\be
\label{h20}
{\cal H}=c_0 + {{\cal Q}\over r^2}
-\ft1{16}\mu^2 \left(\rho^2 + 6L^2\log r\right),
\ee
where $\rho^2=r^2+x^2+y^2+w^2+z^2$.  This generalizes the solution of
\cite{bain}. In the notation of section \ref{sec:n2} we find that
$a=-b=1/2$. For $\H=0$ we have a horizon at $r=0$. However, for the general
expression for $\H$ (\ref{h20}) we find $h=-1$ and subsequently the
conditions for the existence of a horizon (\ref{eq:hcabh}) are not
satisfied. In the particular case of ${\cal Q}=0$ the main
contribution comes from the logarithmic term in $\H$ which results in a
regular horizon. 
Thus, for this type of solution we are able to successfully use
the solution generating mechanism to obtain a solution which
asymptotically approaches the BFHP solution and contains the same
horizon as the seed solution.

\subsubsection{Intersecting branes}
\label{sec:d1d5}

As the above examples indicate, the $\H$-deformation can be 
easily applied to many  brane solutions.  For another example, we
consider the intersecting D1-D5 system.  The IIB
fields of interest are the metric, dilaton, RR 3-form and RR 5-form
field strengths.  In the Einstein frame, the appropriate equations of
motion are
\begin{eqnarray}
R_{MN}&=&\ft12\partial_M\phi\partial_N\phi+\ft1{96}F^2_{5\,MN}
+\ft14e^\phi(F^2_{3\,MN}-\ft1{12}g_{MN}F_3^2),\nonumber\\
\square\phi&=&\ft1{12}e^\phi F_3^2,\nonumber\\
dF_3&=&0,\qquad d*(e^\phi F_3)=0,\nonumber\\
dF_5&=&0,\qquad F_5=*F_5,\qquad F_5\wedge F_3=0.
\label{eq:d1d5eom}
\end{eqnarray}
As demonstrated in \cite{bain}, this system admits independently both
D1 $(+,-,0,0)$ and D5 $(+,-,2,2)$ solutions in the plane wave
background. These two solutions can be combined to form a generalization
of the D1-D5 solution having the
form
\begin{eqnarray}
ds^2&=&H_1^{-\fft34}H_5^{-\fft14}[-4dx^+dx^-+{\cal H}dx^{+\,2}]
+H_1^{\fft14}H_5^{-\fft14}d\vec y\,^2+H_1^{\fft14}H_5^{\fft34}d\vec
z\,^2,\nonumber\\
F_5&=&\mu\,dx^+\wedge(dy^1\wedge dy^2\wedge dz^1\wedge dz^2+dy^3\wedge
dy^4\wedge dz^3\wedge dz^4),\nonumber\\
F_3&=&2\,dx^+\wedge dx^-\wedge dH_1^{-1}+\ft16\epsilon_{ijkl}\partial_lH_5
\,dz^i\wedge dz^j\wedge dz^k,\nonumber\\
e^{2\phi}&=&H_1/H_5.
\label{eq:d1d5met}
\end{eqnarray}
Note that, for simplicity of notation, the $4+4$ split of the plane wave
is in the $\{y^1,y^2,z^1,z^2\}$ and $\{y^3,y^4,z^3,z^4\}$ directions, as
indicated explicitly for the 5-form.  Here, $\vec z$ are
overall-transverse coordinates, while $\vec y$ are
relative-transverse coordinates.  The D1 and D5 Harmonic functions
satisfy $\square H_1=0$ and $\square H_5=0$.  For radial symmetry, they
may be written as
\begin{equation}
H_1=1+\fft{Q_1}{r^2},\qquad H_5=1+\fft{Q_5}{r^2},
\end{equation}
where $r^2=\sum z_i^2$ is the overall-transverse radius
\footnote{As expected, by setting ${\cal H}$ to zero and conformally rescaling
the Einstein frame metric (\ref{eq:d1d5met}) with a dilaton factor 
$e^{\phi/2}$, one reobtains the standard D1-D5 metric in the string frame, 
in flat space (\ref{eq:3c5dbh}).}.

For the plane wave background, ${\cal H}$ satisfies the Laplace equation
with source
\begin{equation}
\Bigl[\square+H_5\widetilde{\square}\Bigr]{\cal H}=-\mu^2.
\label{eq:calheqn}
\end{equation}
To obtain the proper plane wave asymptotics, we take
${\cal H}=-\fft1{16}\mu^2\rho^2+\cdots$, where $\rho^2=\sum z_i^2+y_i^2$
contains all eight transverse coordinates.  The appropriate solution to
(\ref{eq:calheqn}) has the form
\begin{equation}
{\cal H}=c_0+\fft{Q_w}{r^2}-\ft1{16}\mu^2(\rho^2-4Q_5\log r),
\end{equation}
where the constant $c_0$ may be absorbed by shifts in $x^-$.
For $\mu^2=0$, this is identical to the well-known 3-charge $D=5$ black
hole, lifted to ten dimensions (compare with (\ref{eq:3c5dbh}), which is
instead given in the string frame), with entropy $S\sim
\sqrt{Q_1Q_5Q_w}$.

Let us investigate what happens to the original $r=0$ horizon of the
seed solution. Here we have $a=-b=1$ and $ h=-1$; this means that the
second set of conditions in (\ref{eq:hcabh}) for the existence
of a horizon is satisfied, and we have succeeded in deforming the solution
in a way that preserves the horizon of the seed solution and changes the
large $|\vec{z}\,|$ asymptotics to that of the BFHP solution. 
Moreover, the horizon is non-singular, just as for the D1-D5 system in flat
space.

\section{Non-extremal deformations}

The examples of the previous section indicate that many brane solutions
initially asymptotic to Minkowski space may be given BFHP plane wave
asymptotics by appropriately turning on a 
null 5-form.  However, in all such examples, we have focused on
extremal brane configurations.  In this section, we turn to the possibility
of constructing non-extremal deformations of the maximally symmetric
BFHP solution without the introduction of additional fields.  Our
starting point is thus the metric and RR 5-form, initially of the form
(\ref{eq:hpp}).

While the metric itself has an $SO(8)$ isometry, the true isometry group
is only $SO(4)\times SO(4)$ because of the 5-form.  This suggests that, when
seeking non-extremal generalizations, one should be content with
preserving only $SO(4)\times SO(4)$ invariance of the metric.
Nevertheless, we proceed first with an ansatz that retains $SO(8)$
invariance of the metric, and subsequently turn to the $SO(4)\times
SO(4)$ case.

\subsection{An $SO(8)$ invariant deformation}

For an $SO(8)$ invariant deformation, we take as generalization of
(\ref{eq:hpp}) a metric of the form
\begin{eqnarray}
ds^2&=&e^{2A}[-4\,dx^+dx^-+{\cal H}dx^{+\,2}]+e^{2B}d\vec z\,^2 ,
\end{eqnarray}
with the 5-form unchanged.  In addition to the original ${\cal H}(\vec z\,)$,
this introduces two additional `blackening' functions $A(\vec z\,)$ and
$B(\vec z\,)$.
Since the 5-form equations of motion are trivially satisfied, we only
need to be concerned with the Einstein equation.  However, in this case,
a non-constant function $A$ eliminates covariant constancy of
$\partial/\partial x^-$, since
$\nabla_i(\partial/\partial x^-)=2e^{2A}\,\Gamma^+{}_{i+}=\partial_i\,e^{2A}$.
Once we have relaxed the covariant constancy of $\partial/\partial x^-$,
additional non-trivial components of the Ricci tensor show up, and there
are more components of the Einstein equation to be solved.
Nevertheless, as demonstrated in the previous section, the `harmonic
function' ${\cal H}$ only arises in the $R_{++}$ component.  Thus the
system essentially separates into a set of source-free equations,
$R_{+-}=0$ and $R_{ij}=0$, as well as an equation for ${\cal H}$ arising
from $R_{++}$.  The former equations may be solved independently of the
structure of the 5-form source.  Thus we may essentially first build up
a vacuum solution and then demonstrate that ${\cal H}$ may be consistently
turned on in such a background.

The Ricci components $R_{+-}$, $R_{++}$ and $R_{ij}$ may be read off
from (\ref{eq:dnsplit}) for $d=8$ and $n=0$.  Specializing to $SO(8)$
symmetry, we take $A$ and $B$ to be  functions of $r=|\vec z\,|$ only.
The resulting vacuum Einstein equations read
\begin{eqnarray}
&&A''+\fft7rA'+2(A'+3B')A'=0,\nonumber\\
&&B''+\fft7rB'+2(A'+3B')(B'+\fft1r)=0,\nonumber\\
&&A''+3B''+A'^2-2A'B'-3B'^2-\fft1r(A'+3B')=0.
\label{eq:so8eins}
\end{eqnarray}
The first two equations are readily solved.  In particular, by defining
$A+3B=\fft12{\cal G}=\fft12\log g$, we obtain the equation
$\partial_r(r^{13}\partial_r  g)=0$, with solution
\begin{equation}
g(r)=1-\left(\fft{r_0}{r}\right)^{12},
\end{equation}
where we have fixed the asymptotics by demanding $g\to1$ as
$r\to\infty$.  Substituting this into the first equation of
(\ref{eq:so8eins}), we find $\partial_r(r^7g\partial_rA)=0$ which admits
a solution $A=\alpha\log f$ where $\alpha$ is a constant and
\begin{equation}
f(r)=\fft{1-(r_0/r)^6}{1+(r_0/r)^6}.
\label{eq:ffunc}
\end{equation}
Finally, the last equation of (\ref{eq:so8eins}) provides a constraint,
$\alpha^2=7/16$.

Putting everything together, this yields a metric of the form
\begin{equation}
ds^2=f^{2\alpha}[-4dx^+dx^-+{\cal H}dx^{+\,2}]+f^{-\fft23\alpha}g^{\fft13}
d\vec z\,^2.
\label{eq:8met}
\end{equation}
Until now, we have been able to completely ignore the harmonic function
${\cal H}(r)$.  This shows up only in the $R_{++}$ equation, which for
(\ref{eq:8met}) simply reads $\widehat{\square} {\cal H}=-\mu^2$,
where $\widehat{\square}$ is given by (\ref{eq:wtsq}), or
\begin{equation}
r^{-7}\partial_r(r^7g\partial_r{\cal H})=-\mu^2.
\label{eq:8heom}
\end{equation}
Just as in the equation for $A$, this is simply the transverse
Laplacian, however this time with a 5-form source.  The solution is
given by
\begin{equation}
{\cal H}(r)=c_0-\fft{\cal Q}{2r_0^6}\log f(r)-\ft1{16}\mu^2\int
\fft{d(r^2)}{g(r)},
\end{equation}
which is the generalization of (\ref{eq:hppheqn}).  Even though one can
solve explicitly for ${\cal H}$, we are mainly interested in its leading
behavior near $r_0$:
\begin{equation}
{\cal H}=-\left(\fft{\cal Q}{2r_0^6}+\fft{\mu^2r_0^2}{96}\right)
\log\fft{3(r-r_0)}{r_0}+\cdots,\qquad r\to r_0.
\end{equation}
A particular case of this solution appeared in \cite{hr2}, where the
$\mu=0$ case was considered ({\it i.e.}~in the absence of 5-form flux).

The function $f$ in Eq.~(\ref{eq:ffunc}) is reminiscent of a blackening
function in isotropic coordinates, and the metric (\ref{eq:8met}) has
a potential horizon at $r=r_0$.  However, for the vacuum solution, it
was shown in \cite{hr2} that $r=r_0$ is in fact a naked curvature
singularity, and not a true horizon.  The addition of the 5-form flux,
while yielding the proper plane wave asymptotics at infinity,
${\cal H}(r)\sim -\fft1{16}\mu^2r^2$, does not change this conclusion.
This may be seen from the analysis of section~\ref{sec:n2} and in particular
(\ref{eq:hcabh}), where $a=\alpha\approx0.6614$ and
$b=\fft16-\fft13\alpha\approx -0.0538$, so that the $a-b\ge 1$ condition
is violated.  Thus we must look elsewhere for a solution with a realized
horizon.

Note that this IIB background may be T-dualized and lifted to provide a
solution to eleven dimensional supergravity in terms of a non-extremal
deformed M2-brane, which generalizes the extremal solution of
\cite{Cvetic:2002hi}.  The result of the lifting is
\begin{eqnarray}
ds_{11}^2&=& {\cal H}^{-\fft23}f^{-\fft43\alpha}
[-f^{4\alpha} dt^2 + dx_1^2+dx_2^2]
+ {\cal H}^{\fft13}g^{\fft13} d\vec z\,^2,\nonumber\\
F_4&=& \mu(dz^1\wedge dz^2\wedge dz^3\wedge dz^4+dz^5\wedge dz^6
\wedge dz^7\wedge dz^8)\nonumber\\
&&- dt\wedge dx^1\wedge dx^2 \wedge d {\cal H}^{-1},
\end{eqnarray}
which indicates that ${\cal H}$ has the form of a M2-brane harmonic function.
It may be shown that this eleven-dimensional solution also has a naked
singularity at $r=r_0$.

\subsection{An $SO(4)\times SO(4)$ deformation}

Having explicitly demonstrated that the $SO(8)$ symmetric solution leads
to a naked singularity, we now consider the possibility of introducing
a regular horizon by deforming (\ref{eq:hpp}) in an $SO(4)\times SO(4)$
symmetric manner.  This situation is very similar to that of the large
black hole in $AdS_5\times S^5$.  We therefore proceed with a metric
ansatz
\begin{equation}
ds^2=e^{2A}[-4\,dx^+dx^-+{\cal H}dx^{+\,2}]
+e^{2B}d\vec z\,^2 +e^{2C}d\vec y\,^2,
\label{eq:44met}
\end{equation}
where $\vec z=\{z^i\}$, $i=1,2,3,4$ and $\vec y=\{y^a\}$, $a=1,2,3,4$.
Based on the large black hole idea, we take the functions $A(\vec z\,)$,
$B(\vec z\,)$ and $C(\vec z\,)$ to be functions of $\vec z$ only, while
we allow the harmonic function ${\cal H}(\vec z,\vec y\,)$ to depend on all
eight transverse coordinates.  The 5-form must be chosen to respect both
self-duality and the Bianchi identity $dF_{5}=0$.  For the above
metric ansatz, the appropriate choice is
\begin{equation}
F_{5}=\mu\,dx^+\wedge
(e^{4B-4C}dz^1\wedge dz^2\wedge dz^3\wedge dz^4+
dy^1\wedge dy^2\wedge dy^3\wedge dy^4).
\end{equation}
In fact, the combination of self-duality and Bianchi is quite restrictive,
and indicates that in general we would be unable to allow $\vec y$
dependence in $B$ or $C$, except in the $SO(8)$ invariant case when
$B=C$.

The non-vanishing components of the Ricci tensor may be obtained from
(\ref{eq:dnsplit}) with $d=4$ and $n=4$.  We again note that ${\cal H}$
only appears in $R_{++}$, where it enters in terms of an appropriate curved
space transverse Laplacian, and that the remaining Einstein equations
are simply $R_{+-}=0$, $R_{ij}=0$ and $R_{ab}=0$.  The latter source-free
equations may be solved as in the previous subsection.  Defining
\begin{equation}
A+B+2C=\ft12{\cal G}=\ft12\log g,\qquad A=\alpha\log f,\qquad C=\beta\log f,
\end{equation}
we find
\begin{equation}
g(r)=1-\left(\fft{r_0}{r}\right)^4,\qquad
f(r)=\fft{1-(r_0/r)^2}{1+(r_0/r)^2},
\label{eq:fgeqn}
\end{equation}
where the constants $\alpha$ and $\beta$ are required to satisfy the
condition
\begin{equation}
\alpha^2+2\alpha\beta+3\beta^2=\ft38.
\label{eq:alpbet}
\end{equation}
This is the equation for an ellipse, and admits solutions for real
$\alpha$ and $\beta$.

Finally, turning to the ${\cal H}$ equation, we find
\begin{equation}
r^{-3}\partial_r(r^3g\partial_r {\cal H})+
f^{-2(\alpha+3\beta)}g^2\widetilde{\square}{\cal H}
=-\mu^2f^{-4(\alpha+3\beta)}g^2.
\label{eq:harme}
\end{equation}
Unlike in the $SO(8)$ invariant case, (\ref{eq:8heom}), here the source
term is no longer simply constant.  In addition, while ${\cal H}$ is a function
of both $r$ and $\tilde r\equiv |\vec y\,|$, these two transverse
coordinates enter asymmetrically in the above.  However, this equation may
be solved by assuming that ${\cal H}$ separates as a sum of functions 
of $r$ and $\tilde r$ respectively.  Although other solutions are possible,
we simply choose ${\cal H}=-\fft1{16}\mu^2(r^2+\tilde r^2)+\overline{\cal
H}(r)$ so as to reproduce the asymptotics at infinity.  
Then Eq.~(\ref{eq:harme}) becomes
\be
r^{-3}\partial_r(r^3g\partial_r\overline{\cal H})=-\ft12\mu^2
f^{-2(\alpha+3\beta)}g^2(2f^{-2(\alpha+3\beta)}-1),
\label{eq:heom}
\ee
which may be integrated twice to obtain $\overline{\cal H}(r)$.  The
large distance asymptotics has the form
\begin{equation}
{\cal H}=c_0+\fft{\cal Q}{r^2}-\ft1{16}\mu^2(r^2+\tilde
r^2)-3(\alpha+3\beta)\mu^2r_0^2\log\fft{r}{r_0}+\cdots,
\end{equation}
while the behavior near $r_0$ is either logarithmic for
$\alpha+3\beta\le\fft34$ or given by a power law, ${\cal H}\sim
(r-r_0)^{-[4(\alpha+3\beta)-3]}$, otherwise.

In this manner, we have obtained a $SO(4)\times SO(4)$ invariant
deformation of the maximally symmetric plane wave of the form
\begin{eqnarray}
ds^2&=&f^{2\alpha}[-4\,dx^+dx^-+{\cal H}dx^{+\,2}]
+f^{-2(\alpha+2\beta)}g\,d\vec z\,^2
+f^{2\beta}d\vec y\,^2,\nonumber\\
F_{5}&=&\mu\,dx^+\wedge (f^{-4(\alpha+3\beta)}g^2
dz^1\wedge dz^2\wedge dz^3\wedge dz^4+
dy^1\wedge dy^2\wedge dy^3\wedge dy^4).\nonumber\\
\label{eq:44soln}
\end{eqnarray}
This solution is similar to the $SO(8)$ invariant example of
(\ref{eq:8met}), in that the harmonic function solving (\ref{eq:heom})
almost entirely decouples from the Einstein equations.  This is a
feature of retaining $\partial/\partial x^-$ as a null Killing
direction, and ensures that the blackening functions $f$ and $g$
essentially have the form as they would for a vacuum solution, even in
the presence of the 5-form flux.

We now investigate whether this solution admits a horizon at $r=r_0$.
Since both $f\sim r-r_0$ and $g\sim r-r_0$ near $r_0$, the leading
expansion of the warp factors is
\be
e^{2A}\sim(r-r_0)^{2\alpha},\qquad
e^{2B}\sim(r-r_0)^{1-2(\alpha+2\beta)}.
\ee
This yields $a=\alpha$ and $b=\fft12-\alpha-2\beta$.  For
$\alpha+3\beta\le\fft34$, so that ${\cal H}$ is logarithmic near $r_0$,
the criteria for a horizon, (\ref{eq:hcabh}), is then
\begin{equation}
\beta<\ft34,\qquad \alpha+\beta\ge\ft34.
\end{equation}
The second inequality is incompatible with the ellipse condition,
(\ref{eq:alpbet}), since
\be
\alpha^2+2\alpha\beta+3\beta^2-\ft38=
(\alpha+\beta)^2+2\beta^2-\ft38\geq2\beta^2+\ft{3}{16}>0.
\ee
On the other hand, if ${\cal H}$ diverges as a power law, the existence
of a horizon requires $\alpha+\beta\geq \frac 34 +\frac {|h|}2$, which is
also incompatible with the ellipse condition, since it amounts to
$\alpha^2+2\alpha\beta +3\beta^2-\frac 38\geq (\frac 34 + \frac {|h|}2)^2
+2\beta^2-\frac 38 >0$.  Thus all pairs $(\alpha, \beta)$ on the ellipse
(\ref{eq:alpbet}) parametrize solutions that have naked singularities at
$r=r_0$.

\section{\rightskip0em plus2em
No horizons for metrics admitting a null Killing vector}
\label{sec:2}

Based on the observations made in the previous section, one may develop
the suspicion that the class of static metrics preserving a null Killing
vector do not allow for a regular horizon.  Here we demonstrate that
this is indeed the case by proving a no-go theorem.  More precisely, we
shall prove that:
\begin{theorem}
In the presence of matter contributing only to $T_{++}$, and with
$\partial/\partial x^+$ asymptotically a timelike Killing vector,
the following metric 
\begin{equation}
ds^2=e^{2A(r)}[-4dx^+dx^-+{\cal H}dx^{+\,2}]+e^{2B(r)}[dr^2+r^2d\Omega_n^2]
+\sum_i e^{2C_i(r)}dy_i^2
\label{eq:pmetans}
\end{equation}
cannot admit a regular $SO(n+1)$ invariant horizon.  
\end{theorem}
Before proceeding to the proof, let us discuss the conditions of the theorem.
The fact that we consider only null matter puts us in the simplest class of
metrics that asymptote to the BFHP solution with no other fields turned on.
Taking $x^+$ as an asymptotic time coordinate (corresponding to
$\partial/\partial x^+$ being a timelike Killing vector)
is a natural consequence of
the requirement that the metric asymptotes to the BFHP solution ($\H\sim
-\fft1{16}\mu^2 \rho^2$). We will comment on the consequences of relaxing
these conditions at the end of this section. 

The proof of the no-go theorem relies on the separation property of
the metric in the presence of a null Killing vector, (\ref{ricci}). This
separation implies that the vacuum Einstein equations for
(\ref{eq:pmetans}) are equivalent to those of a ``conventional''
${\cal H}=0$ metric supplemented by an equation on ${\cal H}$ (resulting
from the $R_{++}$ equation).  Thus we begin by considering
a metric, not in light-cone variables, but rather written as
\begin{equation}
ds^2=-e^{2A(r)}dt^2+e^{2B(r)}[dr^2+r^2d\Omega_n^2]+\sum_a e^{2C_a(r)}
dy_a^2.
\label{eq:genmet}
\end{equation}
The radially-symmetric vacuum Einstein equations are (no sums except
where explicit)
\begin{eqnarray}
0&=R_{tt}&=e^{2A-2B}[A''+\fft{n}rA'+A'(A'+(n-1)B'+\sum C_a')],\nonumber\\
0&=R_{aa}&=-e^{2C_a-2B}[C_a''+\fft{n}rC_a'+C_a'(A'+(n-1)B'+\sum C_j')],
\nonumber\\
0&=R_{\alpha\beta}&=-e^{-2B}g_{\alpha\beta}[B''+\fft{n}rB'+(B'+\fft1r)
(A'+(n-1)B'+\sum C_a')],\nonumber\\
0&=R_{rr}&=-A''-nB''-\sum C_a''-A'^2-\sum C_a'^2 + B'(A'+\sum
C_a'-\fft{n}{r}),\nonumber\\
\end{eqnarray}
where $\alpha$ and $\beta$ denote directions on $S^n$.  As before, the
first three equations admit a solution of the form
\begin{equation}
A+(n-1)B+\sum C_a={\cal G}=\log g,\qquad A=\alpha\log f,\qquad C_a=\gamma_a
\log f,
\end{equation}
where
\begin{equation}
g(r)=1-\left(\fft{r_0}r\right)^{2(n-1)},\qquad
f(r)=\fft{1-(r_0/r)^{n-1}}{1+(r_0/r)^{n-1}}.
\label{eq:genfg}
\end{equation}
We stress that this is the most general solution consistent with the
natural 
boundary conditions at infinity.

Making use of the identity
\begin{equation}
\fft{g'}g\left(\fft{g'}g+\fft{2(n-1)}r\right)=\left(\fft{f'}f\right)^2,
\end{equation}
we find that the $R_{rr}$ equation yields the algebraic constraint
\begin{equation}
(n-1)\left(\alpha^2+\sum\gamma_a^2\right)
+\left(\alpha+\sum\gamma_a\right)^2=n.
\label{eq:elcond}
\end{equation}
Note that all parameters $\{\alpha,\gamma_a\}$ enter symmetrically in the
constraint.  This indicates that singling out the time coordinate in
(\ref{eq:pmetans}) has no effect at the level of the solution.  On the
other hand, since we are interested in identifying a potential horizon,
we find it useful to maintain this $\{\alpha,\gamma_a\}$ split.

Any set of parameters satisfying (\ref{eq:elcond}) provides a solution
to the vacuum Einstein equations.  However, we must further examine
whether the solution admits a regular horizon.  To do so, we consider
timelike radial geodesics in the background (\ref{eq:genmet}).  Since
the metric is time-independent, we immediately identify a conserved
energy $E=e^{2A}\dot t$, where dots refer to derivatives with respect to
the proper time, $\tau$.  Radial geodesics are then parametrized by
solutions of
\begin{equation}
\dot r^2=e^{-2B}(e^{-2A}E^2-1).
\end{equation}
Based on the behavior of $f$ and $g$ given in (\ref{eq:genfg}), we may
expand $e^{2A}$ and $e^{2B}$ near $r_0$.  For $e^{2A}\to0$, the first
term dominates, and we find
\begin{equation}
d\tau\sim E^{-1} (r-r_0)^{\alpha-\fft1{n-1}(\alpha+\sum\gamma_a-1)}dr,
\end{equation}
or
\begin{equation}
dt\sim (r-r_0)^{-\alpha-\fft1{n-1}(\alpha+\sum\gamma_a-1)}dr.
\label{eq:trrel}
\end{equation}
For $r=r_0$ to be a horizon, we demand that the exponent in
(\ref{eq:trrel}) is such that the geodesic cannot reach $r_0$ in finite
coordinate time $t$.  This gives rise to the condition
\begin{equation}
\alpha+\fft1n\sum\gamma_a\ge1,
\label{eq:horcond}
\end{equation}
for the existence of a horizon.

To determine whether solutions to the constraint (\ref{eq:elcond}) are
compatible with the horizon condition, we rewrite (\ref{eq:elcond}) as
\begin{equation}
\Bigl(\alpha+\fft1n\sum\gamma_a\Bigr)^2
+\fft{n-1}{n^2}\Bigl(n\sum\gamma_a^2+(\sum\gamma_a)^2\Bigr)=1.
\end{equation}
However, since the first term is bounded below by unity from
(\ref{eq:horcond}), and since the remaining terms are sums of squares,
we see that the only satisfactory solution has $\alpha=1$ and all
$\gamma_a$'s set to zero.  In this case, the metric (\ref{eq:genmet})
is especially simple, and has the form
\begin{equation}
ds^2=-\left(\fft{1-(r_0/r)^{n-1}}{1+(r_0/r)^{n-1}}\right)^2dt^2
+ \left(1+(r_0/r)^{n-1}\right)^{\fft2{n-1}}[dr^2+r^2d\Omega_n^2]+d\vec
y\,^2.
\label{eq:isosch}
\end{equation}
This is simply an $(n+2)$-dimensional Schwarzschild black hole (written
in isotropic coordinates) tensored with additional flat dimensions.

Thus we have shown that metrics of the form (\ref{eq:genmet}) only admit
$SO(n+1)$ invariant horizons of the Schwarzschild form,
(\ref{eq:isosch}).  This is perhaps not very surprising, as it
corresponds to some limited form of a uniqueness theorem for black
holes%
\footnote{Of course, consideration of black hole uniqueness is
considerably different in dimensions greater than four.  Here we are not
claiming to make a general statement, but rather a statement only in the
context of vacuum solutions of the form (\ref{eq:genmet}).}.
Before returning to the light-cone metric, (\ref{eq:pmetans}), we note
that black $p$-branes of the general form \cite{Duff:1996hp}
\begin{equation}
ds^2=e^{2A}[-fdt^2 + d\vec y\,^2]+e^{2B}[f^{-1}dr^2+r^2d\Omega_n^2]
\label{eq:goodblack}
\end{equation}
fall into this category of blackening, and indeed have regular horizons.

Finally, we are in a position to make contact with the generalized
pp-wave metric of (\ref{eq:pmetans}).  As demonstrated in
section~\ref{sec:n3}, because of the null Killing vector
$\partial/\partial x^-$, the vacuum Einstein equations for
(\ref{eq:pmetans}) are identical to that of (\ref{eq:genmet}) with an
additional longitudinal coordinate (say $y^0$) satisfying the condition
$\gamma_0=\alpha$, supplemented with a transverse Laplace equation on
${\cal H}$
\begin{equation}
\left(f^{\fft2{n-1}(\alpha+\sum_{a\ge0}\gamma_a)}g^{-\fft2{n-1}}\right)
r^{-n}g^{-1}\partial_r(r^ng\partial_r {\cal H})
+ \sum_{b\ge1} f^{-2\gamma_b} \partial_b^2 {\cal H}
=-2f^{-2\alpha}T_{++}
\end{equation}
(where the Einstein equation is $R_{++}=T_{++}$).
This has no effect on the blackening functions $f$ and $g$ of
(\ref{eq:genfg}).

Assuming $\partial/\partial x^+$ to be asymptotically timelike, we now
examine the conditions, (\ref{eq:hcabh}), which determine the existence of
an event horizon for the generalized pp-wave metric (\ref{eq:pmetans}).
Since these conditions depend on the form of $\H$, we consider several
cases in turn.  First, we assume that $\H$ is logarithmic or finite
near the horizon. Then the horizon conditions are simply
\be
\alpha+\beta > -1,\qquad\alpha-\beta\geq 1,
\label{eq:abhc0}
\ee
where $\beta=(1-\alpha-\sum_{a\ge0}\gamma_a)/(n-1)$, in accord with previous
calculations.  Note that the second inequality is the same condition as
(\ref{eq:horcond}), which determined the horizon for the metric
(\ref{eq:genmet}), except that for the metric (\ref{eq:pmetans}) we have
$\gamma_0=\alpha$.  Recalling that the $rr$ Einstein equation imposes a
constraint (\ref{eq:elcond}) which is compatible with the conditions
defining an event horizon only if $\gamma_a=0$ for all $a$'s, we see that
we end up with a contradiction.  Next, we consider the case where $\H$
grows near the horizon ($h<0$); this corresponds to the second case of
(\ref{eq:hcabh}).  However, now the second horizon condition is of the
form $\alpha-\beta\ge1+|h|>1$ (since $|h|>0$), which is even stronger
than (\ref{eq:abhc0}), and hence is never satisfied.  Finally, we note
that if $\H$ blows up negative near $r_0$ (either logarithmically or
as a power) we end up with a repulsive singularity.

As a result, we have demonstrated that all solutions of the form
(\ref{eq:pmetans}) with null matter are necessarily naked singularities
(except for the trivial Minkowski vacuum).  The essential feature here
is that, to maintain a null Killing vector, the `light-cone' directions
$x^+$ and $x^-$ must be `blackened' simultaneously.  This would be similar
to seeking a black string solution of the form
\begin{equation}
ds^2=e^{2A}f[-dt^2+dz^2]+e^{2B}[f^{-1}dr^2+r^2d\Omega_n^2],
\end{equation}
which does not admit regular solutions, in contrast to the appropriate
blackening ansatz of the form (\ref{eq:goodblack}).  As a result, to
obtain a reasonable horizon for the vacuum solution, we must relax the
null Killing vector condition (at least near the horizon).

We now comment on how the theorem may be modified if any of the
conditions is relaxed.  Firstly, the introduction of additional non-null
matter would modify the Einstein equations that determine $A$, $B$ and
$C_i$ in a way that depends on the concrete type of matter involved.
Thus we are unable to make any general statements without explicitly
specifying the non-null matter content.  As indicated in the next section,
turning on additional matter is not always enough for the evasion of the
no-go theorem.

Secondly, the identification of an asymptotic time coordinate is another
crucial condition.  So long as $\H<0$ in the asymptotic region,
$\partial/\partial x^+$ is asymptotically a timelike Killing vector, and
the above analysis applies.   However, for $\partial/\partial x^+$ null or
spacelike, one must instead resort to the horizon criteria (\ref{eq:hcabh2}).
For $h\ge 0$ the condition is precisely what we had previously,
(\ref{eq:abhc0}), and the no-go theorem can be repeated along the same
lines. On the other hand, for $h<0$ the horizon condition becomes
$\alpha-\beta\ge1-|h|$.  In particular, it is a weaker condition than
the one used in proving the no-go theorem.  In this sense, provided 
$\partial/\partial x^+$ is not timelike, turning on $\H$ improves
the chances of having a regular event horizon.  As indicated by the
examples considered in section \ref{sec:n2}, solutions are easily found
that have a regular horizon.

Finally, it is interesting to note that generically the equation for $\H$
contains a generalized Laplacian that allows for a homogeneous part in
the solution of $\H$. Namely, there is usually the possibility of adding
a term such that $\widehat{\square}_d \H=0$. The solution to this part
contains, in principle, $h<0$ and  therefore creates room for having an
event horizon whenever the asymptotic time has a linear component along
$x^-$.

\section{Turning on additional sources}
\label{sec:3}

Since deformations of the plane wave background with any splitting and
global symmetry $SO(m)\times SO(n)\times SO(p)\times\cdots$ fall into the
category of ansatze dealt with by the precedent no-go theorem, one might
hope that turning on additional fluxes ({\it i.e.}~matter) would help
in evading the previous negative outcome.  Perhaps the most natural
starting point is a string-type deformation of the plane wave, {\it
i.e.}~an object extended in the $(+,-)$ directions.  However, based on
the general structure of the Einstein equations, it is also likely that any
D-brane solution would admit a non-extremal deformation.  Thus, after
considering the string-like case, we turn to ansatze which generalize
the extremal D-brane solutions in the maximal supersymmetric plane wave
background of the type IIB supergravity.

\subsection{Sourcing a string}

The string ansatz which we analyze has a partial smearing in two
transverse directions, with the metric ansatz corresponding to a $2+6$ split
\be
ds^2=e^{2A(r)}[-4dx^+dx^-+{\cal H}dx^{+\,2}]+e^{2B(r)}[dr^2+
r^2 d\Omega_5 {}^2]+e^{2C(r)}(dy_1^2 +dy_2^2),
\ee
and with the NS-NS 2-form and dilaton field given by
\be
H_{+-r}=F'(r),\qquad\square \phi=-\ft1{12}e^{-\phi}H^2.
\ee
The stress tensor components determined by these fields are
\bea
T_{+-}&=&\ft3{16}e^{-2A-2B}e^{-\phi}F'(r)^2,\nonumber\\
T_{ab}&=&\ft1{32}\delta_{ab}e^{-4A-2B+2C}e^{-\phi}F'(r)^2,\nonumber\\
T_{ij}&=&\ft1{32}e^{-4A}e^{-\phi}F'(r)^2(\delta_{ij}-4\hat x_i\hat x_j)
+\ft12 \partial_i \phi\partial_j\phi.
\eea
So instead of solving for vacuum Einstein equations for all components
with the notable exception of the $++$ component, we now have sources for
each Einstein equation.  However, in taking appropriate linear combinations
of the $(+-), (ab)$ and the ${ij}$ components, the sources cancel and we
can solve for
\be
A+C+2B=\ft12{\cal G}=\ft12 \log g,\qquad A+3C=\gamma \log f,
\label{eq:abceqn}
\ee
where $f, g$ are
\be
g(r)=1-\left(\frac{r_0}{r}\right)^8,\qquad
f(r)=\frac{1-(r_0/r)^4}{1+(r_0/r)^4},
\ee
and $\gamma$ is a constant of integration.

To proceed, we note that the NS-NS and dilaton field equations yield
\bea
F'&=&-8e^{\phi+4A}\frac{L^4}{r^5 g},\nonumber\\
\phi'' + \phi'(\frac{g'}g +\frac 5r)&=&8e^{\phi+4A}\frac{L^8}{r^{10}g^2},
\label{eq:fpeqn}
\eea
where $L$ is an integration constant related to the NS-NS 3-form
(string) charge.  Finally, the remaining Einstein equations read
\bea
&&A''+A'(\frac{g'}g+\frac5r)=6e^{\phi+4A}\frac{L^8}{r^{10}g^2},\nonumber\\
&&A''+C''+2B''-\frac
1r(A'+C'+2B')+A'^2+C'^2-2B'^2-2A'B'-2B'C'\nonumber\\
&&\kern10cm =-\frac 14 \phi'^2+4e^{\phi+4A}\frac{L^8}{r^{10}g^2}.\kern1cm
\label{eq:ee2}
\eea
Combining the first equation of (\ref{eq:ee2}) with the second equation
of (\ref{eq:fpeqn}) allows us to solve for $\phi$ in terms of $A$.  We
obtain
\be
\frac{\phi}4-\frac A3=\beta \log f,
\label{eq:aphieqn}
\ee
where $\beta$ is an (as yet) undetermined constant.  Now, using
(\ref{eq:abceqn}) and (\ref{eq:aphieqn}), as well as the identity
\begin{equation}
\biggl(\fft{g'}g\biggr)^2+\fft8r\fft{g'}g=\biggl(\fft{f'}f\biggr)^2,
\end{equation}
the second equation of (\ref{eq:ee2}) can be rewritten as an equation
for $A$ only:
\be
\biggl[\biggl(e^{-8\hat A/3}\biggr)'\biggr]{}^2
-\lambda^2 \biggl(e^{-8\hat A/3}\biggr)^2 \biggl(\frac {f'}{f}\biggr)^2
=16\frac {L^8}{r^{10} g^2},
\label{str}
\ee
where
\begin{equation}
\hat A=A+ \frac {3\beta}4\log f,
\end{equation}
and
\begin{equation}
\lambda^2=\frac{15-4\gamma^2-72\beta^2}6.
\label{eq:s2lameqn}
\end{equation}
This equation admits a solution
\begin{equation}
e^{-8\hat A/3}=H\equiv\frac {L^4}{2\lambda r_0^4}\sinh(c-{\lambda}\log f),
\label{eq:h2def}
\end{equation}
where $c$ is an integration constant (chosen so that $H\to1$ as
$r\to\infty$) and we assume that $\lambda^2$ is positive.

Putting the above results together, we find the string solution to have
the form
\begin{eqnarray}
&\displaystyle
ds^2=H^{-\fft34}f^{-\fft32\beta}[-4dx^+dx^-+{\cal H}dx^{+\,2}]
+H^{\fft14}f^{\fft12\beta-\fft13\gamma}[g^{\fft12}(dr^2+r^2d\Omega_5^2)
+f^{\gamma}(dy_1^2+dy_2^2)],&\nonumber\\
&\displaystyle
H_{+-r}=-8H^{-2}\fft{L^4}{r^5g},\qquad
e^{2\phi}=H^{-1}f^{6\beta}.&
\end{eqnarray}
Here $H$, defined in (\ref{eq:h2def}), is identified with the
fundamental string, since $H\approx 1+L^4/r^4$ in the limit
$r_0\to0$.  Note, however, that $H$ is modified from the standard
F1 harmonic function for finite $r_0$, and develops a singularity:
\begin{equation}
H\sim (r-r_0)^{-|\lambda|}\quad\hbox{as}\quad r\to r_0.
\end{equation}
As a result, the behavior of the metric near $r_0$ is given by
\begin{equation}
e^{2A}\sim (r-r_0)^{\fft34|\lambda|-\fft32\beta},\qquad
e^{2B}\sim (r-r_0)^{-\fft14|\lambda|+\fft12\beta-\fft13\gamma+\fft12}.
\end{equation}
In order to investigate the horizon properties of this solution, we need
one extra piece of information, namely the behavior of ${\cal H}$.
{}From the (++) Einstein equation we obtain
\be
\widehat{\square}{\cal H} + e^{-2C}\widetilde{\square} {\cal H}=
-\mu^2 e^{-2A-6B-2C},
\label{calh1}
\ee
where $\widehat{\square}$ is defined in (\ref{eq:wtsq}) and
$\widetilde{\square}=\partial_{y_1}^2+\partial_{y_2}^2$.
Again, writing ${\cal H}=-\frac1{16}\mu^2(r^2+y_1^2+y_2^2)+\overline{\cal
H}(r)$ to guarantee proper asymptotic behavior, the above equation becomes
\be
r^{-5}\partial_r(r^5 g \partial_r \overline{\cal H})=
\ft14\mu^2(g^{\fft32} f^{-\gamma}-g).
\label{eq:f1calhsoln}
\ee

In general, solutions to (\ref{eq:f1calhsoln}) will develop a logarithmic
singularity at $r_0$ (so long as $\gamma<\fft52$, which is always
satisfied for $\lambda^2$ positive).  Thus, from (\ref{eq:hcabh}), the
criteria for the existence of a regular horizon is
\begin{equation}
|\lambda|>2\beta+\ft23\gamma-5,\qquad
|\lambda|\ge2\beta-\ft13\gamma+\ft52,
\end{equation}
where $\lambda$ is given in (\ref{eq:s2lameqn}).  It is easy to see
that, while the first inequality is always satisfied, the second one is
never satisfied for real $\beta$ and $\gamma$.  Thus we must conclude
that this solution is ill behaved whenever $r_0\ne0$.  In fact, $r=r_0$
is a naked singularity, as the curvature invariant $R_{\mu\nu\rho\sigma}
R^{\mu\nu\rho\sigma}$ blows up at $r_0$.

\subsection{Deforming the extremal D-branes}

In section~\ref{sec:3.2.2}, we have examined extremal D3 branes in a plane
wave background.  Here we consider non-extremal deformations of both the D3
$(+,-,1,1)$ and D3 $(+,-,2,0)$ branes. Even though some of these solutions were
naked singularities one might hope that a nonzero Hawking temperature
horizon cloaks the original singularity as it was the case for the
nonzero temperature generalization of the 
Klebanov-Tseytlin solution \cite{kttemp}. Of course, the separation 
property of the Ricci tensor (\ref{ricci}) into background and plane wave
pieces allows non-extremal deformations of arbitrary D branes as well.
However we are content with presenting the following examples.

\subsubsection{The D3 $(+,-,1,1)$ branes}

For the D3 $(+,-,1,1)$ brane, the supergravity background in question is
given in (\ref{eq:d3pm11}).  A natural generalization of this solution
is given by
\begin{eqnarray}
ds^2&=& e^{2A}[-4dx^+\, dx^- + {\cal H} dx^{+\,2}]
+ e^{2B} d\vec z_6^{\,2} +e^{2C} ( dx^2 + dy^2 ),\nonumber\\
F_5^{(0)}&=&2\cdot(1+ *)\,dx^+ \wedge dx^- \wedge dx\wedge dy \wedge d
e^{2D},\nonumber\\
F_5^{(1)}&=& \mu\, dx^+\wedge (dx\wedge dz^1\wedge dz^2\wedge dz^3
+ dy \wedge dz^4\wedge dz^5\wedge dz^6).
\end{eqnarray}
This allows for both additional blackening functions in the metric and a
possible charge modification in the 5-form.  The energy momentum tensor
for this configuration, $T_{MN}=\fft1{96}F^2_{MN}$, decomposes as a sum
of two independent components, $T_{MN}=T_{MN}^{(0)}+T_{MN}^{(1)}$ where
\begin{eqnarray}
T_{\mu\nu}^{(0)}&=&-\ft14g_{\mu\nu}e^{-4A-2B-4C+4D}(\partial_iD)^2\quad
(\mu,\nu=+,-,x,y),\nonumber\\
T_{ij}^{(0)}&=&\ft14e^{-4A-4C+4D}[\delta_{ij}(\partial_kD)^2-2\partial_iD
\partial_jD],
\end{eqnarray}
is a standard D3-brane source, while
\begin{equation}
T_{++}^{(1)}=\ft12\mu^2e^{-6B-2C}
\end{equation}
corresponds to the plane wave deformation.  The vanishing of a possible
cross-term between $F_5^{(0)}$ and $F_5^{(1)}$ is not {\it a priori}
guaranteed, but is crucial for the existence of the solution \cite{bain}.

Since the Ricci tensor of (\ref{eq:rmnsplit}) similarly splits into two
components, the equations of motion separate into: $i$) the standard
3-brane equations $R_{MN}^{(0)}=T_{MN}^{(0)}$ (for functions $A$, $B$,
$C$ and $D$), and $ii$) a Poisson equation for ${\cal H}$
\begin{equation}
\widehat{\square}{\cal H}+e^{-2C}\widetilde{\square}{\cal
H}=-\mu^2e^{-2A-6B-2C},
\end{equation}
where $\widehat{\square}$ is defined in (\ref{eq:wtsq}) and
$\widetilde{\square}=\partial_x^2+\partial_y^2$.  Note that this is
identical to (\ref{calh1}), which defined ${\cal H}$ for the F1 string
solution. This is expected, of course, since ${\cal H}$
characterizes the plane wave part of the metric, and obeys an equation
sensitive only to the splitting of the transverse directions, in
this case $2+6$.

{}From the standard
3-brane equations, we can easily obtain certain linear combinations
of the warp factors in terms of the blackening functions $f(r)$ and
$g(r)$:
\bea
&&A+2B+C=\ft12{\cal G}=\ft12\log g,\qquad A-C=\gamma\log f,\\
&&g(r)=1-\left({r_0\over r}\right)^8,\qquad
f(r)=\frac{1-(r_0/r)^4}{1+(r_0/r)^4}.
\eea
Furthermore, the only non-trivial Bianchi identity
\be
\left(r^5 e^{-2A+4B-2C} (e^{2D})'\right)'=0,
\ee
can be integrated once to yield the D3-brane potential in terms of the
other warp factors.
Finally, just as in the string solution, the remaining Einstein equation
can be written, after a little algebra, as
\be
\biggl[\biggl(e^{-4\hat A}\biggr)'\biggr]^2
-\lambda^2\biggl(e^{-4\hat A}\biggr)^2
\biggl(\frac{f'}f\biggr)^2 = 16\frac{L^8}{r^{10}g^2},
\ee
where
\be
\hat A=A-\frac\gamma 2\log f,
\ee
and
\begin{equation}
\lambda^2=\frac{5-4\gamma^2}2.
\end{equation}
This equation admits a solution of the form
\begin{equation}
e^{-4\hat A}=H\equiv\frac {L^4}{2\lambda r_0^4}
\sinh(c-\lambda\log f),
\end{equation}
where $c$ is an integration constant, chosen so that $H\to1$ as
$r\to\infty$.  The D3 $(+,-,1,1)$ metric is thus given by
\begin{equation}
ds^2=H^{-\fft12}f^\gamma[-4dx^+dx^-+{\cal H}dx^{+\,2}
+f^{-2\gamma} (dx^2+dy^2)]+H^{\fft12}g^{\fft12} [dr^2+r^2d\Omega_5^2].
\end{equation}
As expected, our solution reduces to the extremal case of (\ref{eq:d3pm11})
if we set $r_0=0$ ({\it i.e.}~$f=g=1$) \cite{bain}.  Note that, in this
extremal limit, $H=1+L^4/r^4$ is simply the D3 harmonic function, while for
finite $r_0$, we have instead $H\sim(r-r_0)^{-|\lambda|}$ as $r\to r_0$.

The horizon analysis then proceeds as in the string case.  By writing
${\cal H}=-\fft1{16}\mu^2(r^2+x^2+y^2)+\overline{\cal H}(r)$, we obtain
\begin{equation}
r^{-5}\partial_r(r^5g\partial_r\overline{\cal H})=\ft14\mu^2(g^{\fft32}
f^\gamma H-g),
\end{equation}
which develops at most a logarithmic singularity at $r_0$.  As a result,
the conditions for a regular horizon are
\begin{equation}
\gamma> -\ft52,\qquad |\lambda|+\gamma\ge\ft52.
\end{equation}
Again, the second inequality is never solved, and this solution
corresponds to a naked singularity. 


\subsubsection{The D3 $(+,-,2,0)$ branes}

The D3 $(+,-,2,0)$ branes of (\ref{eq:d3pm20}) span the directions
$(+, -, 2,0)=(+,-,x,y)$, such that their
worldvolume lies partly in only one of the 4+4 dimensional transverse
spaces. They are supported by an additional 5-form flux $F_{+- xy r}(r)$
where $r$ is the radius of the 4 dimensional transverse space which does
not include the $(x,y)$ directions, and are partially smeared since all
metric components with the exception of $g_{++}$ depend only on $r$. The
D3 $(+,-,2,0)$ branes preserve 1/4 of the linearly realized
supersymmetries.  An ansatz which captures these features and
generalizes (\ref{eq:d3pm20}) is
\bea
ds^2&=&e^{2A(r)}[-4dx^+dx^-+{\cal H}(r,x,y,w,z) dx^{+\,2}]\nonumber\\
&&+e^{2B(r)}[dr^2+ r^2 d\Omega_3 {}^2]
+e^{2C(r)}(dx^2 +dy^2)+ e^{2D(r)}(dw^2+ dz^2),\nonumber\\
F_{5}^{(0)}&=&2\cdot(1+*)\,dx^+ \wedge dx^- \wedge dx \wedge dy \wedge
de^{2E(r)},\nonumber\\
F_{5}^{(1)}&=&\mu\,dx^+\wedge\left(dx\wedge dy \wedge dw\wedge dz+
e^{4B-2C-2D} r^3~dr\wedge d\Omega_3\right).
\eea
As in the D3 $(+,-,1,1)$ case, the Einstein equations separate cleanly
into a set of standard D3-brane equations and an additional Laplace
equation for ${\cal H}$.

By taking appropriate linear combinations of Einstein equations such that
the source terms cancel, we learn that
\be
A+B+C+D=\ft12{\cal G}=\ft12\log g(r),\qquad
A+D=\delta \log f(r),\qquad A-C=\gamma\log f(r),
\ee
where, as we have become accustomed to, the functions $g(r)$ and $f(r)$ have
the same universal expression sensitive
only to the dimensionality and particular splitting of the transverse space
\be
g(r)=1-\left(\frac{r_0}{r}\right)^4,\qquad
f(r)=\frac{1-(r_0/r)^2}{1+(r_0/r)^2}.
\ee
The only non-trivial Bianchi identity yields
\be
\left(r^3 e^{-2A+2B-2C+2D}(e^{2E})' \right)'=0.
\ee
Integrating this equation once, one can substitute it into the remaining
Einstein equation:
\be
\biggl[\biggl(e^{-4\hat A}\biggr)'\biggr]^2-
\lambda^2\biggl(e^{-4\hat A}\biggr)^2\biggl(\frac {f'}f\biggr)^2
=4\frac {L^4}{r^6g^2},
\ee
where
\begin{equation}
\hat A= A-\frac \gamma 2 \log f,
\end{equation}
and
\begin{equation}
\lambda^2={3-8\delta^2+8\delta\gamma-4\gamma^2}.
\label{eq:d320elip}
\end{equation}
The solution is again of the form
\begin{equation}
e^{-4\hat A}=H\equiv\frac{L^2}{2\lambda r_0^2} \sinh(c- \lambda\log f),
\end{equation}
with $c$ chosen so that $H\to1$ as $r\to\infty$.  As a result, the
non-extremal D3 $(+,-,2,0)$ metric has the form
\begin{eqnarray}
ds^2&=&H^{-\fft12}f^\gamma[-4dx^+dx^-+{\cal H}dx^{+\,2}
+f^{-2\gamma}(dx^2+dy^2)]\nonumber\\
&&+H^{\fft12}f^{\gamma-2\delta}[g(dr^2+r^2d\Omega_3^2)
+f^{-2\gamma+4\delta}(dw^2+dz^2)],
\end{eqnarray}
and reduces to the extremal case (\ref{eq:d3pm20}) with $H=1+L^2/r^2$,
when $r_0=0$.  
The plane wave function ${\cal H}$ can be solved from the (++) Einstein
equation
\be
\widehat{\square}{\cal H}+e^{-2C} \square_{x,y} {\cal H}
+e^{-2D} \square_{w,z} {\cal H}=- \mu^2 e^{-2A-4C-4D}.
\ee
The resulting behavior at $r_0$ is generically logarithmic, unless
$|\lambda|+6\delta-4\gamma>3$, whereupon ${\cal H}$ would instead be
dominated by a power law singularity, ${\cal H}\sim
(r-r_0)^{-[|\lambda|+6\delta-4\gamma-3]}$.

Extracting the behavior of the warp factors near $r_0$, we find the
horizon conditions
\begin{equation}
\gamma-\delta>-\ft32,\qquad -|\lambda|+2\delta\ge3,
\end{equation}
(provided ${\cal H}$ is at most logarithmic at $r_0$).  These conditions,
in conjunction with (\ref{eq:d320elip}), do not admit real solutions for
$\gamma$ and $\delta$.  Note, furthermore, that the case where ${\cal
H}$ develops a power law behavior imposes even stronger inequalities,
and the solution is again singular.

\subsubsection{The D1-D5 system}

Finally, we note that the D1-D5 system examined in section
\ref{sec:d1d5} also admits a non-extremal generalization.  The
techniques for obtaining such a solution are similar to that for the
cases considered above.  As a result, we will be brief, and will only
highlight the salient features of the solution.

The equations of motion
governing the D1-D5 system were previously given in (\ref{eq:d1d5eom}),
and admitted the extremal solution (\ref{eq:d1d5met}).  This solution
may be generalized by introducing additional blackening functions in the
metric
\begin{equation}
ds^2=H_1^{-\fft34}H_5^{-\fft14}f^{2\alpha}[-4dx^+dx^-+{\cal H}dx^{+\,2}]
+H_1^{\fft14}H_5^{-\fft14}f^{2\gamma}d\vec y\,^2
+H_1^{\fft14}H_5^{\fft34}f^{2\beta}g\,d\vec z\,^2,
\end{equation}
where
\begin{equation}
g(r)=1-\left(\fft{r_0}r\right)^4,\qquad
f(r)=\fft{1-(r_0/r)^2}{1+(r_0/r)^2},
\end{equation}
with $r=|\vec z\,|^2$.
The constants $\alpha$, $\beta$ and $\gamma$ are then chosen to satisfy
$2\alpha+2\beta+4\gamma=0$, so that $\cal G$ defined after
(\ref{eq:dnsplit}) takes on the canonical form ${\cal G}=\log g$.

As is well known, the D1 and D5 charges may become renormalized in the
non-extremal case.  Thus, we can no longer assume the RR 3-form to
be given in terms of the same $H_1$ and $H_5$ functions that appear in
the metric.  We find the proper generalization of $F_3$ given
in (\ref{eq:d1d5met}) to be
\begin{equation}
F_3=-2H_1^{-2}f^{2\delta}g^{-1}dx^+\wedge dx^-\wedge
d\left(\fft{Q_1}{r^2}\right)
+\ft16\epsilon_{ijkl}\partial_l\left(\fft{Q_5}{r^2}\right)dz^i\wedge
dz^j\wedge dz^k,
\end{equation}
where $\delta$ is a constant to be determined.  This ansatz along with
the $F_3$ equation of motion then determines the dilaton to be
\begin{equation}
e^{\phi}=H_1^{\fft12}H_5^{-\fft12}f^{4\alpha-2\delta}.
\end{equation}
The remaining equations to be solved are the Einstein and dilaton
equations of motion.

Solving the remaining equations yields the explicit form of $H_1$ and
$H_5$:
\begin{eqnarray}
H_1&=&\fft{Q_1}{2\lambda_1r_0^2}\sinh(c_1-\lambda_1\log f),\nonumber\\
H_5&=&\fft{Q_5}{2\lambda_5r_0^2}\sinh(c_5-\lambda_5\log f),
\label{eq:h1h5eqn}
\end{eqnarray}
as well as the constraints $\alpha=\beta$, $\gamma=0$ and
\begin{equation}
\lambda_1^2+\lambda_5^2+32\alpha^2=3.
\label{eq:d1d5econ}
\end{equation}
The constants $c_1$ and $c_5$ are chosen so that $H_1, H_5\to1$ as
$r\to\infty$.  In this case, the ``harmonic functions'' have asymptotic
behavior
\begin{equation}
H_1=1+\fft{\sqrt{Q_1^2+4\lambda_1^2r_0^4}}{r^2}+\cdots,\qquad
H_5=1+\fft{\sqrt{Q_5^2+4\lambda_5^2r_0^4}}{r^2}+\cdots,
\end{equation}
indicating explicitly the effect of charge renormalization.  On the
other hand, near $r_0$, the functions behave as
\begin{equation}
H_1\sim(r-r_0)^{-|\lambda_1|},\qquad
H_5\sim(r-r_0)^{-|\lambda_5|}.
\end{equation}

Collecting the above results, we find the non-extremal generalization of
the D1-D5 solution solution to be
\begin{eqnarray}
ds^2&=&H_1^{-\fft34}H_5^{-\fft14}f^{2\alpha}[-4dx^+dx^-+{\cal H}dx^{+\,2}]
+H_1^{\fft14}H_5^{-\fft14}f^{-2\alpha}d\vec y\,^2
+H_1^{\fft14}H_5^{\fft34}f^{2\alpha}g\,d\vec z\,^2 ,\nonumber\\
F_5&=&\mu\,dx^+\wedge(dy^1\wedge dy^2\wedge dz^1\wedge dz^2+dy^3\wedge
dy^4\wedge dz^3\wedge dz^4),\nonumber\\
F_3&=&-2H_1^{-2}g^{-1}dx^+\wedge dx^-\wedge
d\left(\fft{Q_1}{r^2}\right)
+\ft16\epsilon_{ijkl}\partial_l\left(\fft{Q_5}{r^2}\right)dz^i\wedge
dz^j\wedge dz^k,\nonumber\\
e^{2\phi}&=&H_1H_5^{-1}f^{8\alpha},
\end{eqnarray}
where $H_1$ and $H_5$ are given by (\ref{eq:h1h5eqn}), and $\H$ satisfies
\begin{equation}
r^{-3}\partial_r(r^3g\partial_r\H)+H_5f^{4\alpha}g^2\widetilde{\square}\H
=-\mu^2,
\end{equation}
which limits to (\ref{eq:calheqn}) when $r_0\to0$.  Again, proper
asymptotics may be obtained by setting $\H=-\fft1{16}\mu^2(r^2+\vec
y\,^2)+\overline{\H}(r)$ where
\begin{equation}
r^{-3}\partial_r(r^3g\partial_r\overline{\H})=\ft12\mu^2(H_5f^{4\alpha}g^2-1).
\end{equation}

Turning to the issue of horizons, we find that
\begin{equation}
e^{2A}\sim(r-r_0)^{\fft34|\lambda_1|+\fft14|\lambda_5|+2\alpha},\qquad
e^{2B}\sim(r-r_0)^{-\fft14|\lambda_1|-\fft34|\lambda_5|+2\alpha+1},
\end{equation}
in the language of section \ref{sec:n2}.  Thus the conditions for the
existence of a horizon are
\begin{equation}
|\lambda_1|-|\lambda_5|+8\alpha >-6,\qquad
|\lambda_1|+|\lambda_5|\ge3.
\label{eq:d1d5hcon}
\end{equation}
The second inequality is incompatible with the ellipsoidal constraint
(\ref{eq:d1d5econ}).  Hence the solution is a naked singularity.  Note
that in (\ref{eq:d1d5hcon}) we have implicitly assumed that $\H$ develops
at most a logarithmic singularity.  Since no horizon is possible in this
case, neither can it be possible for the case where $\H$ has a power law
behavior.  Thus, once more, in attempting to non-extremalize well
behaved solutions, we instead end up with a naked singularity at $r_0$.  

While these string, D3 and D1-D5 examples have not been exhaustive, they
nevertheless suggest that the essence of the no-go theorem proven in the
previous section continues to apply, even in the presence of additional
matter sources.  The common feature in all these cases is of course the
null Killing vector, whose preservation requires a simultaneous blackening
of two longitudinal directions.  This results in an `overblackened'
solution, which is necessarily a naked singularity.  Thus it is likely
that one must give up the null Killing vector in order to avoid the
appearance of such singularities.  This will be examined in the
following section.

\section{Relaxing the null Killing vector condition}

Until now, we have focused on metrics admitting a null Killing vector
as the simplest generalization allowing a horizon and capturing some of
the features of the BFHP solution. However, as we have seen, both the
no-go theorem of
section~\ref{sec:2} (in the absence of sources) and the examples of
section~\ref{sec:3} (with additional sources turned on) suggest that the
null Killing vector condition may still be too restrictive, at least for
obtaining a non-extremal solution with regular horizon.  This difficulty
arises in a way since the null Killing vector, being $\partial/\partial x^-$,
is not obviously related to the null direction in $(t,r)$ that would exist
at the horizon of {\it e.g.}~a Schwarzschild black hole.  As a result,
this incompatibility of symmetries apparently prevents the existence of
a regular horizon. To proceed, it is clear that we must relax the null
Killing vector condition.  However, to maintain some relation to the BFHP
solution, we must ensure that proper boundary conditions are satisfied so
that the geometry may at least asymptotically be regarded as the BFHP
solution. 

Let us further consider the general structure of the solution we are
seeking. The general solution would naturally contain two scales: $\mu$,
giving the asymptotic strength of the RR 5-form, and $r_0$, essentially
the Schwarzschild radius or non-extremality scale.  Finding an explicit
solution for arbitrary values of $r_0$ and $\mu$ has proven to be a
difficult task.  However, we may consider various limiting cases where
we have a rough intuition of what the solution should look like. For
example, in the limiting case $r_0\mu\ll1$, the two main scales are
widely separated.  This limit is essentially that of a small black hole
in the plane wave background%
\footnote{The situation is similar to the small black hole in
AdS. Although the existence of this small black hole is conceptually
clear, finding an explicit solution has not been possible so far
\cite{Horowitz:2000kx}. The main obstruction is the same symmetry
considerations we face here.}.

With this separation of scales, if we focus on the region $r\gg\mu^{-1}$,
the details of the black hole ought to be irrelevant, and the metric
should end up being asymptotic to the plane wave (\ref{eq:hpp}).  On the
other hand, in the region where $r\ll\mu^{-1}$, the black hole geometry
dominates, and the 5-form flux supporting the background may be considered
as a perturbation to the Schwarzschild black hole.  This is precisely
the limit we consider here. We thus will examine the consistency of
this solution by exploring the possibility of perturbatively turning on
a 5-form in a Schwarzschild-like metric.  Namely, we begin by considering
a black string in IIB theory which is the natural starting point (due
to the symmetries) for the plane wave.

We find it convenient to work in isotropic coordinates.  The black
string in ten dimensions may be written as
\begin{equation}
ds^2=-f^2dt^2+dy^2+h^{\fft23}d\vec z\,^2,
\end{equation}
where
\begin{equation}
f=\fft{1-(r_0/r)^6}{1+(r_0/r)^6},\qquad h=1+\left(\fft{r_0}r\right)^6.
\label{eq:fhform}
\end{equation}
Here, $r=|\vec z\,|$ is an eight-dimensional transverse coordinate.  By
defining $y=x^+-2x^-$ and $t=2x^-$, the black string metric may be
written in the equivalent form
\begin{equation}
ds^2=-4dx^+dx^-+dx^{+\,2}+4(1-f^2)dx^{-\,2}+h^{\fft23}d\vec z\,^2.
\label{eq:bhb}
\end{equation}
In the limit $r\to\infty$ we see that $f\to1$ and therefore $g_{--}\to 0$.
Hence $\partial/\partial x^-$, which is a Killing vector, becomes
asymptotically null.  This allows us to make connection with the
plane wave at infinity.

We now seek to turn on a RR 5-form in the background
(\ref{eq:bhb}).  We first note that {\it no solution is possible that
maintains the accidental $SO(8)$ symmetry}.  This is easily seen by
writing the most general 5-form ansatz compatible with both this
symmetry and self-duality as
\begin{eqnarray}
F_{5}&=&(f_1\,dx^+-2(f_1+f_2f)dx^-)
\wedge dz^1\wedge dz^2\wedge dz^3\wedge dz^4\nonumber\\
&&+(f_2\,dx^+-2(f_2+f_1f)dx^-)\wedge dz^5\wedge dz^6\wedge dz^7\wedge dz^8,
\end{eqnarray}
where $f_1(r)$ and $f_2(r)$ are arbitrary functions of $r$.  The Bianchi
identity $dF_{5}=0$ then demands that $f_1$, $f_2$, $f_1f$ and
$f_2f$ are all constant.  For $r_0=0$ ($f=1$), this allows the solution
(\ref{eq:hpp}).  However, for $r_0\ne0$, this is only compatible with the
complete vanishing of $F_{5}$.  We have thus demonstrated that the
non-extremal solution cannot maintain the accidental $SO(8)$ invariance
of the metric.  This should not come as a surprise, because once we
abandon the null Killing vector $\partial/\partial x^-$, we essentially
return to the Freund-Rubin case, where the 5-form is responsible for the
negative and positive curvatures of AdS$_5\times S^5$.

Of course, the $SO(8)$ symmetry of (\ref{eq:hpp}) was never a true
symmetry in the first place.  So there is no difficulty in giving it up.
However, we expect to still retain the $SO(4)\times SO(4)$ symmetry.  To
do so, we write the transverse metric as
\begin{equation}
d\vec z\,^2=dr^2+r^2[d\theta^2+\sin^2\theta d\Omega_3^2+\cos^2\theta
d\widetilde\Omega_3^2].
\end{equation}
The most general 5-form allowed by symmetry and self-duality is now
considerably more complicated.  Following \cite{Horowitz:2000kx}, we
determine
\begin{eqnarray}
F_{5}&=&g_4r^3\sin^4\theta\,dx^+\wedge dr\wedge d\Omega_3
+g_3r^4\sin^3\theta\cos\theta\,dx^+\wedge d\theta\wedge
d\Omega_3\nonumber\\
&&+g_2r^3\cos^4\theta\,dx^+\wedge dr\wedge d\widetilde\Omega_3
-g_1r^4\cos^3\theta\sin\theta\,dx^+\wedge d\theta\wedge
d\widetilde\Omega_3\nonumber\\
&&+2(g_1f-g_4)r^3\sin^4\theta\,dx^-\wedge dr\wedge d\Omega_3
+2(g_2f-g_3)r^4\sin^3\theta\cos\theta\,dx^-\wedge d\theta\wedge
d\Omega_3\nonumber\\
&&+2(g_3f-g_2)r^3\cos^4\theta\,dx^-\wedge dr\wedge d\widetilde\Omega_3
-2(g_4f-g_1)r^4\cos^3\theta\sin\theta\,dx^-\wedge d\theta\wedge
d\widetilde\Omega_3,\nonumber\\
\label{eq:5fans}
\end{eqnarray}
where $g_i(r,\theta)$, $i=1,2,3,4$ are in principle functions of both $r$ and
$\theta$.  After suitable manipulation, the Bianchi identities read
\begin{eqnarray}
(-\cot\theta\partial_\theta+4)g_2&=&(r\partial_r+4)g_1,\nonumber\\
(\tan\theta\partial_\theta+4)g_1&=&(r\partial_r+r\partial_r\log f+4)g_2,
\nonumber\\
(\tan\theta\partial_\theta+4)g_4&=&(r\partial_r+4)g_3,\nonumber\\
(-\cot\theta\partial_\theta+4)g_3&=&(r\partial_r+r\partial_r\log
f+4)g_4.
\label{eq:biag}
\end{eqnarray}
Note that this splits into two pairs of equations, one for $g_1,g_2$ and
another identical set for $g_3,g_4$.

In the extremal case ($f=1$), the Bianchi identities (which are the
5-form equations of motion) have a trivial solution $g_i=\mu$,
$i=1,2,3,4$.  This yields precisely the 5-form of (\ref{eq:hpp}).  More
generally, we wish to retain this asymptotic behavior, $g_i\to\mu$ as
$f\to1$, but otherwise solve (\ref{eq:biag}) for $f$ of the form
(\ref{eq:fhform}).  Concentrating on $g_1,g_2$, the first two lines of
(\ref{eq:biag}) may be differentiated further to obtain the second order
equations
\begin{eqnarray}
&&\!\!\!\!
\left(\fft1{r^9f}\partial_r r^9f\partial_r+\fft1{r^2\sin^5\theta\cos^3\theta}
\partial_\theta\sin^5\theta\cos^3\theta\partial_\theta+\fft4r\partial_r\log
f\right)g_1(r,\theta)=0,\nonumber\\
&&\!\!\!\!
\left(\fft1{r^9f}\partial_r r^9f\partial_r+\fft1{r^2\sin^3\theta\cos^5\theta}
\partial_\theta\sin^3\theta\cos^5\theta\partial_\theta+\fft5r\partial_r\log
f+\partial_r^2\log f\right)g_2(r,\theta)=0.\nonumber\\
\end{eqnarray}
These are simply the second order equations of motion for the
5-form.  In general, solutions do exist compatible with the asymptotic
condition $g_i\to\mu$ as $r\to\infty$.  However, for simplicity, we now
assume that $g_i(r)$ are independent of $\theta$.  In this case, we only
need to solve
\begin{equation}
\left(\fft1{r^9f}\partial_rr^9f\partial_r+\fft4r\partial_r\log
f\right)g_1(r)=0.
\label{eq:g1eqn}
\end{equation}
Then $g_2$ may be obtained from the first equation of (\ref{eq:biag}),
namely $g_2(r)=(1+\fft14r\partial_r)g_1(r)$.

\begin{figure}[t]
\epsfxsize10cm
\begin{center}
\leavevmode\epsfbox{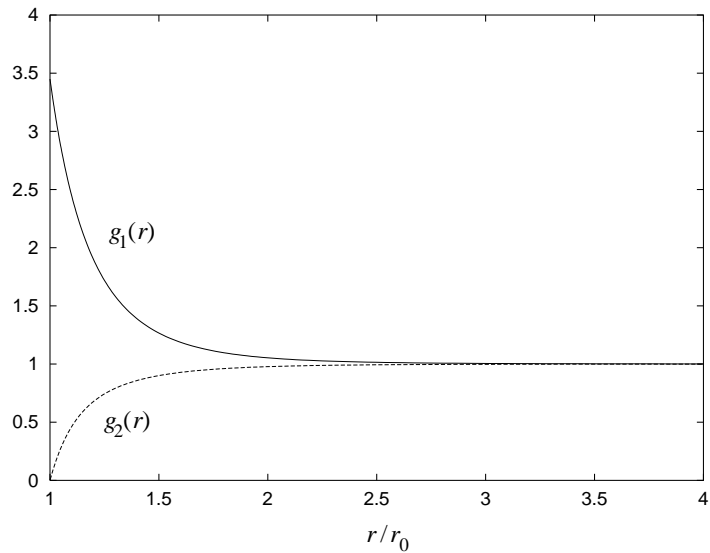}
\end{center}
\caption{The functions $g_1$ and $g_2$ normalized such that $g_i\to1$ as
$r\to\infty$.  Note that $g_2$ vanishes at the horizon.}
\label{fig:1}
\end{figure}

Since (\ref{eq:g1eqn}) is a linear second order equation, it admits two
solutions.  It is easy to see that as $r\to\infty$ the solutions have
behavior $g_1\sim 1$ and $g_1\sim r^{-8}$.  The former yields the proper
asymptotics, while the latter falls off faster than the blackening
function and is subdominant.  On the other hand, the horizon behavior of
$g_1$ is governed by the limit $f\to0$ and may be obtained by expanding
(\ref{eq:g1eqn}) for $r\to r_0$.  Near the horizon, we find either
$g_1\sim 1$ or $g_1\sim\log(r-r_0)$, so that one solution remains
finite, while the second blows up.  Keeping only the solution that
remains finite at the horizon, this gives a unique $\theta$-independent
solution for the $g_i$'s, up to overall normalization.  The numerical
solution to (\ref{eq:g1eqn}) which satisfies such boundary conditions is
shown in Fig.~\ref{fig:1}.  As a result, this demonstrates that the 5-form
with the desired asymptotics for connecting to the plane wave of
(\ref{eq:hpp}) may be consistently turned on in the black string 
background.

Although we do not compute the back-reaction from turning on the 5-form,
it is nevertheless instructive to examine the stress-energy source.  For
the 5-form of (\ref{eq:5fans}), we find
\begin{eqnarray}
F^2_{++}&=&48h^{-\fft83}G_+^2,\nonumber\\
F^2_{+-}&=&96h^{-\fft83}(g_1g_4f-G_+^2),\nonumber\\
F^2_{--}&=&192h^{-\fft83}(G_+^2(1+f^2)-2g_1g_2f),
\end{eqnarray}
for the longitudinal directions, and
\begin{eqnarray}
F^2_{rr}&=&48h^{-2}G_-^2,\nonumber\\
F^2_{\theta\theta}&=&-48h^{-2}r^2G_-^2,\nonumber\\
F^2_{\alpha\beta}&=&48h^{-2}r^2\sin^2\theta g_{\alpha\beta}
G_-^2(\sin^2\theta-\cos^2\theta),\nonumber\\
F^2_{\tilde\alpha\tilde\beta}&=&-48h^{-2}r^2\cos^2\theta
g_{\tilde\alpha\tilde\beta}
G_-^2(\sin^2\theta-\cos^2\theta),
\end{eqnarray}
for the transverse ones.  Here, for convenience, we have defined
$G_\pm^2\equiv\ft12(g_4^2\pm g_1^2)$.  The asymptotic behavior is
guaranteed as $G_-^2\to0$ as $r\to\infty$, while all quantities remain
finite at the horizon.  This suggests that the back-reaction is under
control.  Furthermore, we see explicitly that the stress energy source
in the transverse directions will break the $SO(8)$ symmetry of the
metric (\ref{eq:bhb}), once back-reaction is taken into account.
While we have not obtained a complete non-linear solution, this
analysis in the regime $r_0\mu\ll1$ provides strong evidence that an
appropriate non-extremal gravitational background {\it does exist}. 

It is worth mentioning that there are other approaches for understanding
the phase diagram of asymptotically BFHP solutions. A natural attempt,
complementary to the one presented in this section, 
would be to consider the motion of a particle in the plane wave background
and the subsequent effects on the geometry of increasing the mass of the
particle%
\footnote{We thank A. Hashimoto and E. Gimon for discussions of
this possibility.}.
This approach has been recently discussed in \cite{linear}. 

The complete understanding of phase transitions (like the Hawking-Page
phase transition) and its gauge theory interpretation along the lines of
\cite{witten} requires knowledge of the solution for arbitrary values of
$r_0$ and $\mu$. It seems likely at this point that such a phase diagram
might be more intricate than the corresponding one for asymptotically
AdS spaces, in which case finding its complete structure would require
a mixture of approaches. We believe, however, that our arguments for the
existence of a solution with a well-defined Schwarzschild-type horizon
adds a crucial element to the understanding of the ultimate phase diagram.

\begin{center}
\large{Acknowledgments}
\end{center}

We would like to thank A. Hashimoto, I. Klebanov, F. Larsen,
J. Maldacena, D. Marolf and R. Roiban.
We are especially grateful to E. Gimon for a very fruitful
conversation.  J.T.L was supported in part by DOE Grant
DE-FG02-95ER40899. L.A.P.Z. was supported by a grant in aid from the
Funds for Natural Sciences at I.A.S. The work of D.V. is   
supported by DOE grant DE-FG02-91ER40671.




\begin{thebibliography}{99}

\bibitem{ads}
J.~Maldacena,
{\sl The large $N$ limit of superconformal field theories and supergravity},
Adv.\ Theor.\ Math.\ Phys.\  {\bf 2} (1998) 231 [hep-th/9711200].\\
%
S.S.~Gubser, I.R.~Klebanov and A.M.~Polyakov,
{\sl Gauge theory correlators from non-critical string theory},
Phys.\ Lett.\  {\bf B428} (1998) 105 [hep-th/9802109].\\
%
E.~Witten,
{\sl Anti-de Sitter space and holography},
Adv.\ Theor.\ Math.\ Phys.\  {\bf 2} (1998) 253 [hep-th/9802150].

\bibitem{hawking}
S.~W.~Hawking and D.~N.~Page,
{\sl Thermodynamics Of Black Holes In Anti-de Sitter Space},
Commun.\ Math.\ Phys.\  {\bf 87} (1983) 577.

\bibitem{witten}
E.~Witten,
{\sl Anti-de Sitter space, thermal phase transition, and confinement in
gauge theories},
Adv.\ Theor.\ Math.\ Phys.\  {\bf 2} (1998) 505 [hep-th/9803131].

\bibitem{clifford}
A.~Chamblin, R.~Emparan, C.~V.~Johnson and R.~C.~Myers,
{\sl Charged AdS black holes and catastrophic holography},
Phys.\ Rev.\ D {\bf 60} (1999) 064018 [hep-th/9902170].\\
%
A.~Chamblin, R.~Emparan, C.~V.~Johnson and R.~C.~Myers,
{\sl Large $N$ phases, gravitational instantons and the nuts and bolts of AdS
holography},
Phys.\ Rev.\ D {\bf 59} (1999) 064010 [hep-th/9808177].

\bibitem{bmn}
D.~Berenstein, J.~M.~Maldacena and H.~Nastase,
{\sl Strings in flat space and pp waves from ${\cal N} = 4$ super Yang Mills},
JHEP {\bf 0204} (2002) 013 [hep-th/0202021].

\bibitem{blau}
M.~Blau, J.~Figueroa-O'Farrill, C.~Hull and G.~Papadopoulos,
{\sl A new maximally supersymmetric background of IIB superstring theory},
JHEP {\bf 0201}, 047 (2002) [hep-th/0110242].

\bibitem{kg}
J.~Kowalski-Glikman,
{\sl Vacuum States In Supersymmetric Kaluza-Klein Theory},
Phys.\ Lett.\ B {\bf 134}, 194 (1984).





\bibitem{metsaev}
R.~R.~Metsaev,
{\sl Type IIB Green-Schwarz superstring in plane wave Ramond-Ramond
background},
Nucl.\ Phys.\ B {\bf 625} (2002) 70 [hep-th/0112044].

\bibitem{metse}
R.~R.~Metsaev and A.~A.~Tseytlin,
{\sl Exactly solvable model of superstring in plane wave Ramond-Ramond
background},
Phys.\ Rev.\ D {\bf 65} (2002) 126004 [hep-th/0202109].\\
%
J.G.~Russo and A.A.~Tseytlin
{\sl On solvable models of type 2B superstring in NS-NS and R-R wave
backgrounds},
JHEP {\bf 0204} (2002) 021 [hep-th/0202179].

\bibitem{thermal}
L.~A.~Pando Zayas and D.~Vaman,
{\sl Strings in RR plane wave background at finite temperature},
hep-th/0208066.\\
%
B.~R.~Greene, K.~Schalm and G.~Shiu,
{\sl On the Hagedorn behaviour of pp-wave strings and N = 4 SYM theory at
finite R-charge density},
hep-th/0208163.\\
%
Y.~Sugawara,
{\sl Thermal amplitudes in DLCQ superstrings on pp-waves},
hep-th/0209145.\\
%
R.~C.~Brower, D.~A.~Lowe and C.~I.~Tan,
{\sl Hagedorn transition for strings on pp-waves and tori with chemical
potentials},
hep-th/0211201.

\bibitem{cobi}
L.~A.~Pando Zayas and J.~Sonnenschein,
{\sl On Penrose limits and gauge theories},
JHEP {\bf 0205} (2002) 010 [hep-th/0202186].

\bibitem{donald}
D.~Marolf and L.~A.~Pando~Zayas,
{\sl On the singularity structure and stability of plane waves},
hep-th/0210309.

\bibitem{hr1}
V.~E.~Hubeny and M.~Rangamani,
{\sl No horizons in pp-waves},
JHEP {\bf 0211}, 021 (2002) [hep-th/0210234].

\bibitem{hr2}
V.~E.~Hubeny and M.~Rangamani,
{\sl Generating asymptotically plane wave spacetimes},
hep-th/0211206.

\bibitem{garfinkle}
D.~Garfinkle,
{\sl Traveling Waves In Strongly Gravitating Cosmic Strings},
Phys.\ Rev.\ D {\bf 41} (1990) 1112.\\
%
D.~Garfinkle and T.~Vachaspati,
{\sl Cosmic String Traveling Waves},
Phys.\ Rev.\ D {\bf 42} (1990) 1960.\\
%
D.~Garfinkle,
{\sl Black String Traveling Waves},
Phys.\ Rev.\ D {\bf 46} (1992) 4286 [gr-qc/9209002].

\bibitem{myers}
N.~Kaloper, R.~C.~Myers and H.~Roussel,
{\sl Wavy strings: Black or bright?},
Phys.\ Rev.\ D {\bf 55} (1997) 7625 [hep-th/9612248].

\bibitem{wald}
R.~M.~Wald,
{\it General Relativity},
The University of Chicago Press, 1984.


\bibitem{horatiu}
D.~Berenstein and H.~Nastase,
{\sl On lightcone string field theory from super Yang-Mills and holography},
hep-th/0205048.

\bibitem{donross}
D.~Marolf and S.~F.~Ross,
{\sl Plane waves: To infinity and beyond!},
Class.\ Quant.\ Grav.\  {\bf 19} (2002) 6289 [hep-th/0208197].

\bibitem{hr3}
V.~E.~Hubeny and M.~Rangamani,
{\sl Causal structures of pp-waves},
hep-th/0211195.


\bibitem{hayward}
S.~A.~Hayward,
{\sl General Laws Of Black Hole Dynamics},
Phys.\ Rev.\ D {\bf 49}, 6467 (1994) [gr-qc/9309004].


\bibitem{5dstring}
M.~Cvetic and D.~Youm,
{\sl Dyonic BPS saturated black holes of heterotic string on a six
torus},
Phys.\ Rev.\ D {\bf 53} (1996) 584 [hep-th/9507090].

\bibitem{stringmore}
M.~Cvetic and A.~A.~Tseytlin,
{\sl General class of BPS saturated dyonic black holes as exact superstring
solutions},
Phys.\ Lett.\ B {\bf 366} (1996) 95 [hep-th/9510097].\\
%
M.~Cvetic and A.~A.~Tseytlin,
{\sl Solitonic strings and BPS saturated dyonic black holes},
Phys.\ Rev.\ D {\bf 53} (1996) 5619 [Erratum-ibid.\ D {\bf 55} (1997) 3907]
[hep-th/9512031].

\bibitem{Cvetic:2002hi}
M.~Cvetic, H.~Lu and C.~N.~Pope,
{\sl Penrose limits, pp-waves and deformed M2-branes},
hep-th/0203082.

\bibitem{bain}
P.~Bain, P.~Meessen and M.~Zamaklar,
{\sl Supergravity solutions for D-branes in Hpp-wave backgrounds},
hep-th/0205106.

\bibitem{a1}
M.~Alishahiha and A.~Kumar,
{\sl D-brane solutions from new isometries of pp-waves},
Phys.\ Lett.\ B {\bf 542} (2002) 130 [hep-th/0205134].

\bibitem{kostas}
K.~Skenderis and M.~Taylor,
{\sl Branes in AdS and pp-wave spacetimes},
JHEP {\bf 0206} (2002) 025 [hep-th/0204054].

\bibitem{matthias}
M.~R.~Gaberdiel and M.~B.~Green,
{\sl The D-instanton and other supersymmetric D-branes in IIB plane-wave
string theory},
hep-th/0211122.

\bibitem{bain2}
P.~Bain, K.~Peeters and M.~Zamaklar,
{\sl D-branes in a plane wave from covariant open strings},
hep-th/0208038.

\bibitem{kostas2}
K.~Skenderis and M.~Taylor,
{\sl Open strings in the plane wave background. I: Quantization and
symmetries},
hep-th/0211011.

\bibitem{matthias2}
O.~Bergman, M.~R.~Gaberdiel and M.~B.~Green,
{\sl D-brane interactions in type IIB plane-wave background},
hep-th/0205183.


\bibitem{Duff:1996hp}
M.~J.~Duff, H.~Lu and C.~N.~Pope,
{\sl The black branes of M-theory},
Phys.\ Lett.\ B {\bf 382}, 73 (1996) [hep-th/9604052].

\bibitem{kttemp}
A.~Buchel,
{\sl Finite temperature resolution of the Klebanov-Tseytlin singularity},
Nucl.\ Phys.\ B {\bf 600} (2001) 219 [hep-th/0011146].\\
%
A.~Buchel, C.~P.~Herzog, I.~R.~Klebanov, L.~A.~Pando Zayas and A.~A.~Tseytlin,
{\sl Non-extremal gravity duals for fractional D3-branes on the conifold},
JHEP {\bf 0104} (2001) 033 [hep-th/0102105].\\
%
S.~S.~Gubser, C.~P.~Herzog, I.~R.~Klebanov and A.~A.~Tseytlin,
{\sl Restoration of chiral symmetry: A supergravity perspective},
JHEP {\bf 0105} (2001) 028 [hep-th/0102172].

\bibitem{Horowitz:2000kx}
G.~T.~Horowitz and V.~E.~Hubeny,
{\sl Note on small black holes in $AdS_p\times S^q$},
JHEP {\bf 0006}, 031 (2000) [hep-th/0005288].

\bibitem{linear}
M.~Li,
{\sl Correspondence principle in a pp-wave background},
Nucl.\ Phys.\ B {\bf 638} (2002) 155 [hep-th/0205043].\\
%
M.~Li,
{\sl PP-wave black holes and the matrix model},
hep-th/0212345.


\end{thebibliography}
\end{document}